\documentclass[twocolumn,twocolappendix]{aastex631}
\graphicspath{{figures}}

\usepackage{afterpage}
\usepackage{lipsum}
\usepackage{enumitem}
\usepackage{physics}
\usepackage{multirow}
\usepackage{textgreek}
\usepackage{booktabs}
\usepackage{xfrac}
\hypersetup{pdfnewwindow=true,
     colorlinks=true,
     linkcolor=blue, 
     citecolor=blue,
	 final=true}
	 
\usepackage{etoolbox }
\makeatletter
\patchcmd{\NAT@citex}
  {\@citea\NAT@hyper@{%
     \NAT@nmfmt{\NAT@nm}%
     \hyper@natlinkbreak{\NAT@aysep\NAT@spacechar}{\@citeb\@extra@b@citeb}%
     \NAT@date}}
  {\@citea\NAT@nmfmt{\NAT@nm}%
   \NAT@aysep\NAT@spacechar\NAT@hyper@{\NAT@date}}{}{}
\patchcmd{\NAT@citex}
  {\@citea\NAT@hyper@{%
     \NAT@nmfmt{\NAT@nm}%
     \hyper@natlinkbreak{\NAT@spacechar\NAT@@open\if*#1*\else#1\NAT@spacechar\fi}%
       {\@citeb\@extra@b@citeb}%
     \NAT@date}}
  {\@citea\NAT@nmfmt{\NAT@nm}%
   \NAT@spacechar\NAT@@open\if*#1*\else#1\NAT@spacechar\fi\NAT@hyper@{\NAT@date}}
  {}{}
\makeatother

\newcommand{\JWST}{\textit{JWST}}
\newcommand{\HST}{\textit{HST}}

\newcommand{\oiii}{[O\,\textsc{iii}]}
\newcommand{\ha}{H$\alpha$}

\newcommand{\um}{\textmu m}
\newcommand{\Msun}{{\rm M}_\odot}

\begin{document}

%\title{\large Whether dust-reddened AGN or compact starbursts, \JWST's \\ ``little red dots'' represent a key stage of early galaxy formation}
%Compact galaxies or reddened AGN? Either way

%\title{\large JWST ``little red dots'': a large population of luminous, dust-reddened AGN or massive, compact starbursts? Insights from the COSMOS-Web survey}

%\title{\large JWST ``little red dots'': a population of luminous, dust-reddened AGN or massive, compact starbursts in the early universe?}

%Characterizing the elusive population of compact red 
%COSMOS-Web: 
%COSMOS-Web: A large sample of compact red objects 

% COSMOS-Web: Whether reddened quasars or compact starbursts, JWST's Little Red Dots imply a significant galaxy/SMBH assembly 
%challenge early galaxy formation 

% COSMOS-Web: Hundreds of Little Red Dots reveal 

%\title{\large COSMOS-Web: Whether compact galaxies or red quasars, JWST's \\ little red dots dominate early galaxy formation} 
\title{\large COSMOS-Web: The over-abundance and physical nature of ``little red dots''---Implications for early galaxy and SMBH assembly}

%The origin of compact, extremely red objects discovered by JWST: dusty AGN or nuclear starbursts?

\shortauthors{Akins et al.}
\shorttitle{}

\suppressAffiliations 
\correspondingauthor{Hollis B. Akins} 
\email{hollis.akins@gmail.com}

\author[0000-0003-3596-8794]{Hollis B. Akins}
\altaffiliation{NSF Graduate Research Fellow}
\affiliation{Department of Astronomy, The University of Texas at Austin, 2515 Speedway Blvd Stop C1400, Austin, TX 78712, USA}

\author[0000-0002-0930-6466]{Caitlin M. Casey}
\affiliation{Department of Astronomy, The University of Texas at Austin, 2515 Speedway Blvd Stop C1400, Austin, TX 78712, USA}
\affiliation{Cosmic Dawn Center (DAWN), Denmark}
\author[0000-0003-3216-7190]{Erini Lambrides}
\affiliation{NASA Goddard Space Flight Center, 8800 Greenbelt Rd, Greenbelt, MD 20771, USA}
\altaffiliation{NPP Fellow}

\author[0000-0001-9610-7950]{Natalie Allen}
\affiliation{Cosmic Dawn Center (DAWN), Denmark}
\affiliation{Niels Bohr Institute, University of Copenhagen, Jagtvej 128, DK-2200, Copenhagen N, Denmark}

\author[0000-0001-6102-9526]{Irham T. Andika}
\affiliation{Technical University of Munich, TUM School of Natural Sciences, Department of Physics, James-Franck-Str. 1, 85748 Garching, Germany}
\affiliation{Max-Planck-Institut für Astrophysik, Karl-Schwarzschild-Str. 1, 85748 Garching, Germany}

\author[0000-0002-0245-6365]{Malte Brinch}
\affil{Cosmic Dawn Center (DAWN), Denmark}
\affiliation{DTU-Space, Technical University of Denmark, Elektrovej 327, DK-2800 Kgs. Lyngby, Denmark}

\author[0000-0002-6184-9097]{Jaclyn B. Champagne}
\affiliation{Steward Observatory, University of Arizona, 933 N Cherry Ave, Tucson, AZ 85721, USA}

\author[0000-0003-3881-1397]{Olivia Cooper}
\altaffiliation{NSF Graduate Research Fellow}
\affiliation{Department of Astronomy, The University of Texas at Austin, 2515 Speedway Blvd Stop C1400, Austin, TX 78712, USA}

\author[0000-0002-0786-7307]{Xuheng Ding}
\affiliation{Kavli Institute for the Physics and Mathematics of the Universe (WPI), The University of Tokyo Institutes for Advanced Study, The University of Tokyo, Kashiwa, Chiba 277-8583, Japan}

\author[0000-0003-4761-2197]{Nicole E. Drakos}
\affiliation{Department of Astronomy and Astrophysics, University of California, Santa Cruz, 1156 High Street, Santa Cruz, CA 95064, USA}

\author[0000-0002-9382-9832]{Andreas Faisst}
\affiliation{Caltech/IPAC, 1200 E. California Blvd., Pasadena, CA 91125, USA}

\author[0000-0001-8519-1130]{Steven L. Finkelstein} % CEERS architect, contributions via discussions 
\affiliation{Department of Astronomy, The University of Texas at Austin, 2515 Speedway Blvd Stop C1400, Austin, TX 78712, USA}

\author[0000-0002-3560-8599]{Maximilien Franco} % COSMOS-Web NIRCam architect, contributions via discussions
\affiliation{Department of Astronomy, The University of Texas at Austin, 2515 Speedway Blvd Stop C1400, Austin, TX 78712, USA}

\author[0000-0002-7530-8857]{Seiji Fujimoto}
\altaffiliation{Hubble Fellow}
\affiliation{Department of Astronomy, The University of Texas at Austin, 2515 Speedway Blvd Stop C1400, Austin, TX 78712, USA}

\author[0000-0002-8008-9871]{Fabrizio Gentile}
\affiliation{University of Bologna, Department of Physics and Astronomy (DIFA), via Gobetti 93/2, I-40129, Bologna, Italy}
\affiliation{INAF -- Osservatorio di Astrofisica e Scienza dello Spazio, via Gobetti 93/3, I-40129, Bologna, Italy}

\author[0000-0001-9885-4589]{Steven Gillman}
\affiliation{Cosmic Dawn Center (DAWN), Denmark}
\affiliation{DTU-Space, Technical University of Denmark, Elektrovej 327, DK-2800 Kgs. Lyngby, Denmark}

\author[0000-0002-0236-919X]{Ghassem Gozaliasl}
\affiliation{Department of Physics, University of Helsinki, P.O. Box 64, FI-00014 Helsinki, Finland}

\author[0000-0003-0129-2079]{Santosh Harish} % COSMOS-Web MIRI architect
\affiliation{Laboratory for Multiwavelength Astrophysics, School of Physics and Astronomy, Rochester Institute of Technology, 84 Lomb Memorial Drive, Rochester, NY 14623, USA}

\author[0000-0003-4073-3236]{Christopher C. Hayward}
\affiliation{Center for Computational Astrophysics, Flatiron Institute, 162 Fifth Avenue, New York, NY 10010, USA}
\author[0000-0002-3301-3321]{Michaela Hirschmann}
\affiliation{Institute for Physics, Laboratory for Galaxy Evolution and Spectral modelling, Ecole Polytechnique Federale de Lausanne, Observatoire de Sauverny, Chemin Pegasi 51, 1290 Versoix, Switzerland}
\affiliation{INAF, Osservatorio Astronomico di Trieste, Via Tiepolo 11, 34131 Trieste, Italy}

\author[0000-0002-7303-4397]{Olivier Ilbert}
\affiliation{Aix Marseille Université, CNRS, CNES, LAM, Marseille, France}

\author[0000-0001-9187-3605]{Jeyhan S. Kartaltepe} % COSMOS-Web architect
\affiliation{Laboratory for Multiwavelength Astrophysics, School of Physics and Astronomy, Rochester Institute of Technology, 84 Lomb Memorial Drive, Rochester, NY 14623, USA}

\author[0000-0002-8360-3880]{Dale D. Kocevski}
\affiliation{Department of Physics and Astronomy, Colby College, Waterville, ME 04901, USA}

\author[0000-0002-6610-2048]{Anton M. Koekemoer}
\affiliation{Space Telescope Science Institute, 3700 San Martin Drive, Baltimore, MD 21218, USA}

\author[0000-0002-5588-9156]{Vasily Kokorev}
\affiliation{Department of Astronomy, The University of Texas at Austin, 2515 Speedway Blvd Stop C1400, Austin, TX 78712, USA}

\author[0000-0001-9773-7479]{Daizhong Liu} 
\affiliation{Max-Planck-Institut für Extraterrestrische Physik (MPE), Giessenbachstr. 1, D-85748 Garching, Germany}

\author[0000-0002-7530-8857]{Arianna S. Long}
\altaffiliation{Hubble Fellow}
\affiliation{Department of Astronomy, The University of Texas at Austin, 2515 Speedway Blvd Stop C1400, Austin, TX 78712, USA}

\author[0000-0002-9489-7765]{Henry Joy McCracken}
\affiliation{Institut d’Astrophysique de Paris, UMR 7095, CNRS, and Sorbonne Université, 98 bis boulevard Arago, F-75014 Paris, France}

\author[0000-0002-6149-8178]{Jed McKinney}
\altaffiliation{Hubble Fellow}
\affiliation{Department of Astronomy, The University of Texas at Austin, 2515 Speedway Blvd Stop C1400, Austin, TX 78712, USA}

\author[0000-0003-2984-6803]{Masafusa Onoue}
\affiliation{Kavli Institute for the Physics and Mathematics of the Universe (WPI), The University of Tokyo Institutes for Advanced Study, The University of Tokyo, Kashiwa, Chiba 277-8583, Japan}
\affiliation{Center for Data-Driven Discovery, Kavli IPMU (WPI), UTIAS, The University of Tokyo, Kashiwa, Chiba 277-8583, Japan}
\affiliation{Kavli Institute for Astronomy and Astrophysics, Peking University, Beijing 100871, People's Republic of China}

\author[0000-0003-2397-0360]{Louise Paquereau}
\affiliation{Institut d’Astrophysique de Paris, UMR 7095, CNRS, and Sorbonne Université, 98 bis boulevard Arago, F-75014 Paris, France}

\author[0000-0002-7093-7355]{Alvio Renzini}
\affiliation{INAF - Osservatorio Astronomico di Padova, Vicolo dell’Osservatorio 5, I-35122 Padova, Italy}

\author[0000-0002-4485-8549]{Jason Rhodes}
\affiliation{Jet Propulsion Laboratory, California Institute of Technology, 4800 Oak Grove Drive, Pasadena, CA 91001, USA}

\author[0000-0002-4271-0364]{Brant E. Robertson}
\affiliation{Department of Astronomy and Astrophysics, University of California, Santa Cruz, 1156 High Street, Santa Cruz, CA 95064, USA}

\author[0000-0002-7087-0701]{Marko Shuntov}
\affiliation{Cosmic Dawn Center (DAWN), Denmark}
\affiliation{Niels Bohr Institute, University of Copenhagen, Jagtvej 128, DK-2200, Copenhagen N, Denmark}

\author[0000-0002-0000-6977]{John D. Silverman}
\affiliation{Department of Astronomy, Graduate School of Science, The University of Tokyo, 7-3-1 Hongo, Bunkyo-ku, Tokyo, 113-0033, Japan}
\affiliation{Kavli Institute for the Physics and Mathematics of the Universe (WPI), The University of Tokyo Institutes for Advanced Study, The University of Tokyo, Kashiwa, Chiba 277-8583, Japan}
\affiliation{Center for Data-Driven Discovery, Kavli IPMU (WPI), UTIAS, The University of Tokyo, Kashiwa, Chiba 277-8583, Japan}
\affiliation{Center for Astrophysical Sciences, Department of Physics \& Astronomy, Johns Hopkins University, Baltimore, MD 21218, USA}

\author[0009-0003-4742-7060]{Takumi S. Tanaka}
\affiliation{Department of Astronomy, Graduate School of Science, The University of Tokyo, 7-3-1 Hongo, Bunkyo-ku, Tokyo, 113-0033, Japan}
\affiliation{Kavli Institute for the Physics and Mathematics of the Universe (WPI), The University of Tokyo Institutes for Advanced Study, The University of Tokyo, Kashiwa, Chiba 277-8583, Japan}
\affiliation{Center for Data-Driven Discovery, Kavli IPMU (WPI), UTIAS, The University of Tokyo, Kashiwa, Chiba 277-8583, Japan}

\author[0000-0003-3631-7176]{Sune Toft}
\affiliation{Cosmic Dawn Center (DAWN), Denmark}
\affiliation{Niels Bohr Institute, University of Copenhagen, Jagtvej 128, DK-2200, Copenhagen N, Denmark}

\author[0000-0002-3683-7297]{Benny Trakhtenbrot}
\affiliation{School of Physics and Astronomy, Tel Aviv University, Tel Aviv 69978, Israel}

\author[0000-0001-6477-4011]{Francesco Valentino}
\affiliation{European Southern Observatory, Karl-Schwarzschild-Strasse 2, D-85748, Garching bei München, Germany}

\author[0000-0002-7051-1100]{Jorge Zavala} % significant contributions on slack 
\affiliation{National Astronomical Observatory of Japan, 2-21-1 Osawa, Mitaka, Tokyo 181-8588, Japan}

\collaboration{41}{\vspace{-25pt}}

%\received{}
%\revised{}
%\accepted{}
\submitjournal{(Affiliations can be found after the references)}

\begin{abstract} 
	\JWST\ has revealed a population of compact and extremely red galaxies at $z\gtrsim4$, which likely host active galactic nuclei (AGN). We present a sample of 434 ``little red dots'' (LRDs), selected from the 0.54 deg$^2$ COSMOS-Web survey. We fit galaxy and AGN SED models to derive redshifts and physical properties; the sample spans $z\sim5$--$9$ after removing brown dwarf contaminants. We consider two extreme physical scenarios: either LRDs are all AGN, and their continuum emission is dominated by the accretion disk, or they are all compact star-forming galaxies, and their continuum is dominated by stars. If LRDs are AGN-dominated, our sample exhibits bolometric luminosities $\sim10^{45-47}$~erg\,s$^{-1}$, spanning the gap between \JWST\ AGN in the literature and bright, rare quasars. We derive a bolometric luminosity function (LF) $\sim100$ times the (UV-selected) quasar LF, implying a non-evolving black hole accretion density of $\sim10^{-4}$~M$_\odot$~yr$^{-1}$~Mpc$^{-3}$ from $z\sim2$--$9$. By contrast, if LRDs are dominated by star formation, we derive stellar masses $\sim10^{8.5-10}\,M_\odot$. MIRI/F770W is key to deriving accurate stellar masses; without it, we derive a mass function inconsistent with $\Lambda$CDM. The median stellar mass profile is broadly consistent with the maximal stellar mass surface densities seen in the nearby universe, though the most massive $\sim50$\% of objects exceed this limit, requiring substantial AGN contribution to the continuum. Nevertheless, stacking all available X-ray, mid-IR, far-IR/sub-mm, and radio data yields non-detections. Whether dominated by dusty AGN, compact star-formation, or both, the high masses/luminosities and remarkable abundance of LRDs implies a dominant mode of early galaxy/SMBH growth.
\end{abstract}

\keywords{\small {\it Unified Astronomy Thesaurus concepts:} \href{http://astrothesaurus.org/uat/594}{Galaxy evolution (594)}; \href{http://astrothesaurus.org/uat/595}{Galaxy formation (595)}; \href{http://astrothesaurus.org/uat/734}{High-redshift galaxies (734)}; \href{http://astrothesaurus.org/uat/847}{Interstellar medium (847)}; \href{http://astrothesaurus.org/uat/1879}{Circumgalactic medium (1879)}}

\section{Introduction}\label{sec:intro}

%
%Over the last 2 decades, wide-area optical/near-IR surveys have dramatically expanded our view of the first billion years of the Universe. 
%The discoveries of UV-luminous galaxies at $z\gtrsim 8$--$10$, ultra-massive dust-obscured galaxies into the epoch of reionization, and hundreds of bright active galactic nuclei at $z > 6$ \citep{fanSurvey2001, matsuokaSubaru2018, inayoshiAssembly2020, fanQuasars2023} have built up a picture of rapid stellar and supermassive black hole (SMBH) mass growth in the early universe.  

%
%\JWST\ has made major advances in our understanding of high-redshift galaxies and active black holes. 
%The remarkable abundance of UV-luminous galaxies at $z\gtrsim 10$ suggests an early start to galaxy formation \citep[e.g.][]{finkelsteinCEERS2022a, harikaneComprehensive2022}. % more: uncover, etc
%Ultra-massive galaxies at $z\sim 5$ have been identified, with implied stellar baryon fractions of $\sim 0.2$--$0.5$ \citep[][Gentile et al.~\textit{in prep}]{xiaoMassive2023}.  
%Numerous high-$z$ AGN have been identified and confirmed via broad Balmer emission lines \citep{harikaneJWST2023,larsonCEERS2023, ublerGANIFS2023, maiolinoJADES2023}, exotic high-ionization lines \citep{maiolinoSmall2024}, compact morphology \citep{onoueCandidate2023,furtakJWST2023}, and X-ray emission \citep{bogdanEvidence2024, gouldingUNCOVER2023}.

The launch of \JWST\ has revealed new classes of objects unseen before; in particular, an abundant population of compact and extremely red objects, so-called ``little red dots'' \citep[LRDs;][]{mattheeLittle2024}.  
These objects are characterized by point-like morphology, red colors from observed-frame $\sim 2$--$5$\,\um, and often exhibit flat or even blue colors from $1$--$2$\,\um. 
The population appears to be ubiquitous from $z\gtrsim 3$ \citep{barroExtremely2024} to $z\sim 9$ \citep{leungNGDEEP2023}, spanning several orders of magnitude in luminosity.
At these redshifts, the distinctive colors imply a blue UV slope with a steep/red optical continuum. 
Initial studies interpreted these objects as high-mass ($\gtrsim 10^{10}\,\Msun$), $z\gtrsim 6$ galaxy candidates \citep[e.g.][]{labbePopulation2023, akinsTwo2023}, with either high equivalent width (EW) emission lines or strong Balmer breaks and likely with significant dust attenuation driving the red optical color. 
Reconciling these high masses with the point-like morphology requires rapid starbursts in compact cores \citep[see e.g.][]{baggenSizes2023}, and implies a high stellar mass density in the early Universe, which for some time appeared to be in tension with expectations for the halo mass function in $\Lambda$CDM \citep{boylan-kolchinStress2023}. 
While more recent mass estimates have brought down the exceedingly large stellar masses, relieving the possible tension with $\Lambda$CDM \citep[e.g.][]{endsleyJWST2022,chworowskyEvidence2023,wangRUBIES2024a}, the sheer abundance of these objects seems to be at odds with a massive galaxy interpretation.

An alternative interpretation of these ``little red dots'' is that they are dominated by emission from an active galactic nucleus (AGN).
Indeed, spectroscopy has been obtained for $\gtrsim 50$ LRDs, in $\sim 80\%$ of cases confirming the high redshifts ($z>5$) and identifying broad Balmer lines \citep[$\gtrsim 2000$ km/s, e.g.][]{kocevskiHidden2023, kokorevUNCOVER2023, killiDeciphering2023, mattheeLittle2024, furtakHigh2024, greeneUNCOVER2024, kocevskiRise2024, wangRUBIES2024}.  
The point-like morphology and broad Balmer emission lines are signatures of unobscured (Type I) AGN; i.e., we have a direct view to the broad-line region (BLR) and the accretion disk, but with foreground dust attenuation either from a dusty interstellar medium (ISM) or polar dust, nuclear to the AGN \citep[see e.g.][]{netzerRevisiting2015, hickoxObscured2018}.  
This configuration is similar to the ``red quasars'' at $1<z<3$ \citep[e.g.][]{urrutiaEvidence2008, glikmanFIRST2MASS2012, banerjiHeavily2012, glikmanWISE2MASS2022}, for which the dust attenuation is thought to originate from the galaxy-scale ISM \citep{banerjiInterstellar2018, templeIII2019}, and which have also been found out to high-$z$ \citep[e.g.][]{katoSubaru2020, fujimotoDusty2022}. 
Although at face value, the LRDs appear to populate the faint end of the AGN UV luminosity function, when accounting for dust obscuration, their intrinsic luminosities can be quite large \citep[$L_{\rm bol} \gtrsim 10^{45}$ erg\,s$^{-1}$; e.g.][]{kokorevCensus2024}.
Consistent with these high luminoisities, many authors have derived large black hole masses relative to the host galaxy mass, implying exotic black hole seed/growth mechanisms \citep[e.g.][]{maiolinoSmall2024, greeneUNCOVER2024, bogdanEvidence2024, kovacsCandidate2024, juodzbalisDormant2024}.
However, the origin of the red continuum emission---whether it is dominated by AGN or host galaxy---remains unclear, but has major implications for our interpretation of these objects as massive galaxies, luminous quasars, or some combination thereof.
%
%In fact, this highlights the key uncertainty remaining with the LRDs: if the derived, dust-corrected bolometric luminosities are to be believed, these objects are remarkably luminous, 

There exists some evidence that the continuum emission from the LRDs is actually dominated by stars. 
For one, multi-band \JWST/MIRI observations from $5$--$25$\,\um\ have shown a remarkably flat SED in the rest-frame mid-IR for a number of LRDs, placing strong upper limits on the contribution from hot dust in the AGN torus, in tension with the implied luminosities of the quasars \citep[][G. Leung et al.~\textit{in prep}]{williamsGalaxies2023, perez-gonzalezWhat2024}.
Moreover, several broad-line objects exhibit clear Balmer breaks in their spectra, implying a dominant contribution from evolved stars in the rest-frame optical, though the precise contribution is degenerate, with stellar mass estimates ranging over 2 orders of magnitude \citep{wangRUBIES2024a}.  
Even still, not all LRDs with spectroscopic coverage exhibit clear broad lines \citep[e.g.][]{barroExtremely2024,greeneUNCOVER2024}, perhaps indicating that the population may not be homogeneous. 
It remains unclear whether the LRDs can be explained entirely by AGN, with unique geometry/dust properties, or also represent a significant population of compact/massive galaxies in the early Universe.

In this work, we present a large sample of 434 $z\gtrsim 5$ LRDs selected from the COSMOS-Web \JWST\ survey. 
With the large sample size, nearly doubling the number of known LRDs \citep{kokorevCensus2024, kocevskiRise2024}, and large on-sky area,  minimizing the impact of cosmic variance, we provide strong constraints on the volume density of this population.
In Section~\ref{sec:data} we describe the COSMOS-Web survey, our data reduction methodology, and ancillary multi-wavelength data.  
In Section~\ref{sec:sample} we describe our sample selection and SED fitting methodology; we fit each object to both galaxy and QSO models to characterize the two alternative scenarios. 
In Section~\ref{sec:agn} we consider the interpretation of the LRDs as AGN; in particular, we derive their contribution to the bolometric luminosity function and black hole accretion density. 
In Section~\ref{sec:sfgs}, we consider the interpretation of the LRDs as instead compact star-forming galaxies; in particular, we present the derived stellar mass function and examine the feasibility of such dense stellar systems. 
In Section~\ref{sec:stacking} we stack the available multi-wavelngth data over the entire sample, providing constraints on the panchromatic SED of the typical LRD. 
In Section~\ref{sec:discussion} we discuss this sample in context with other \JWST\ results, with particular focus on the physical processes that may be responsible for the unique characteristics of the LRD population. 
Throughout this paper, we adopt a \citet{kroupaInitial2002} initial mass function and a cosmology consistent with the \citet{planckcollaborationPlanck2020} results ($H_0=67.66$ km s$^{-1}$ Mpc$^{-1}$, $\Omega_{m,0} = 0.31$). 
All magnitudes are quoted in the AB system \citep{okeAbsolute1974}.

\begin{figure*}
\centering
\includegraphics[width=\linewidth]{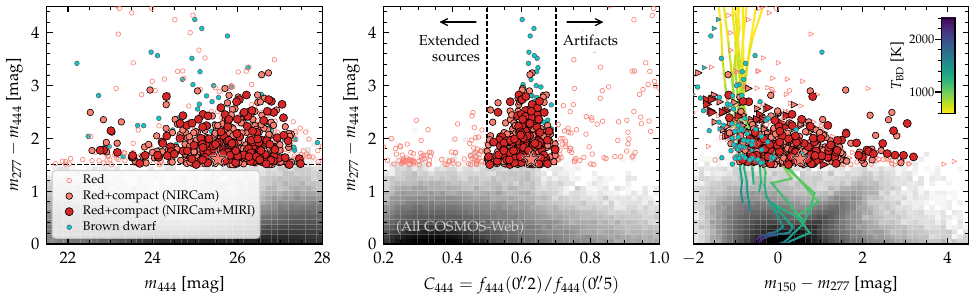}
\caption{Color-magnitude and color-color diagrams illustrating the selection of LRDs in this paper. \textit{Left:} The $m_{277}-m_{444}$ vs.~$m_{444}$ color-magnitude diagram. We select all sources with $S/N_{444} > 12$ (roughly  $m_{444}$ brighter than $27.2$) and $m_{277}-m_{444}>1.5$. Dark red points indicate the subsample which is covered by MIRI imaging, while light red points indicate those with only NIRCam coverage. COS-756434, which is spectroscopically-confirmed at $z=6.99$, is marked with a star. Blue points indicate objects for which a brown dwarf solution is preferred via SED fitting.\textit{Center:} The $m_{277}-m_{444}$ color vs.~compactness $C_{444}$, defined as the ratio of the F444W fluxes in $0\farcs2$ and $0\farcs5$ diameter apertures. Objects with $C_{444}<0.5$ are generally extended, while those with $C_{444}>0.7$ are generally imaging artifacts, i.e.~hot pixels. \textit{Right:} The $m_{277}-m_{444}$ vs.~$m_{150}-m_{277}$ color-color diagram. While we do not select for blue SW colors (either $m_{115}-m_{150}$ or $m_{150}-m_{277}$), as the imaging is too shallow to constrain SW colors at high SNR, we demonstrate here that the NIRCam colors can serve to filter out brown dwarfs, which tend to have bluer SW colors than galaxies at temperatures $T_{\rm BD} < 1200$ K.} \label{fig:selection}	
\end{figure*}

\section{Data}\label{sec:data}

\subsection{COSMOS-Web}

COSMOS-Web is a large \JWST\ Cycle 1 treasury program imaging a contiguous 0.54 deg$^2$ with NIRCam and 0.19 deg$^2$ with MIRI \citep{caseyCOSMOSWeb2023} in the COSMOS field \citep{scovilleCosmic2007}. 
%As of this writing, 6 of the 152 visits have been completed during observations executed in January 2023, constituting a contiguous 77 arcmin2 (∼ 4\% of the overall area), with 8.7 arcmin2 of overlap between the NIRCam and MIRI coverage.
 The COSMOS-Web imaging includes four NIRCam filters (F115W, F150W, F277W, and F444W) at approximate $5\sigma$ depths of $27.4$--$28.2$ AB mag, and one MIRI filter (F770W) at a $5\sigma$ depth of 26 AB mag.
The full details on the NIRCam and MIRI reduction process will be presented in upcoming papers (M.~Franco et al.\,\textit{in prep} and S.~Harish et al.\,\textit{in prep}, respectively) but are briefly described here.
The raw NIRCam imaging was reduced by the JWST Calibration Pipeline version 1.12.1, with the addition of several custom modifications (as has also been done for other JWST studies, e.g.~\citealt{bagleyCEERS2022}) including the subtraction of $1/f$ noise and sky background. 
We use the Calibration Reference Data System (CRDS)\footnote{\url{jwst-crds.stsci.edu}} pmap 1170 which corresponds to NIRCam instrument mapping imap 0273.
The final mosaics are created in Stage 3 of the pipeline with a pixel size of 0\farcs03/pixel. 
Astrometric calibration is conducted via the \JWST/\HST\ alignment tool (JHAT; \citealt{restJWST2023}), with a reference catalog based on an \HST/F814W 0\farcs03/pixel mosaic in the COSMOS field \citep{koekemoerCOSMOS2007} with astrometry tied to Gaia-EDR3 \citep{gaiacollaborationGaia2018}. 
The median offset in RA and Dec between our reference catalog and the NIRCam mosaic is less than 5 mas. 
The MIRI/F770W observations were reduced using version 1.8.4 of the
JWST Calibration pipeline, along with additional steps for background subtraction that was necessary to mitigate instrumental effects. 
The resulting mosaic was resampled onto a common output grid with a pixel scale of 0\farcs03/pixel and aligned with ancillary \HST/F814W imaging of the region.

\subsection{PRIMER-COSMOS}

The Public Release Imaging for Extragalactic Research (PRIMER) survey (P.I.~J.~Dunlop, GO\#1837) is a large Cycle 1 Treasury Program to image two \textit{HST} CANDELS Legacy Fields (COSMOS and UDS) with NIRCam+MIRI \citep{donnanJWST2024}. 
The PRIMER-COSMOS field comprises $\sim 130$ sq.~arcmin and is contained entirely within the COSMOS-Web footprint (see \citealt{caseyCOSMOSWeb2023} for details on the field layout), providing additional depth and filter coverage over $\sim 8\%$ of the field.
PRIMER imaging includes 8 NIRCam bands (adding F090W, F200W, F356W, and F410M to the COSMOS-Web filter set), plus two MIRI bands (adding F1800W). 
We have reduced the PRIMER data following the same procedure as with the COSMOS-Web data.

\subsection{Ancillary Multi-wavelength Data}

While the COSMOS field is covered with substantial ground-based (Subaru and UltraVISTA) and \textit{Spitzer}/IRAC imaging, we do not use these additional photometric data in our SED fitting. 
The lower-resolution of these maps, relative to the \textit{HST}+\textit{JWST} imaging, mandates careful deblending, which can suffer from significant uncertainties.
However, the deep ground-based imaging of the COSMOS field can be useful to constrain photometric redshifts and rule out degenerate low-redshift solutions.
%We therefore inspect cutouts from Subaru Hyper-suprime Cam $grizy$, UltraVISTA $YJHK_s$, and \textit{Spitzer}/IRAC ch. 1, 3, and 4 for each LRD in the sample. 
%Sources with HSC $g+r$ detections are likely lower-redshift contaminants; we remove XX from our sample. 
%We verify that for isolated objects the HSC $grizy$ and UVISTA $YJHK_s$ aperture photometry is consistent with the best-fit SED. 

Even beyond the optical/near-infrared, the COSMOS field is rich with ancillary multiwavelength data in the X-ray, MIR, FIR/submm, and radio.
We make use of \textit{Chandra} X-ray imaging from the COSMOS Legacy Survey \citep{civanoChandra2016}, \textit{Spitzer}/MIPS 24+70 µm imaging from S-COSMOS \citep{sandersSCOSMOS2007}, \textit{Herschel}/PACS+SPIRE imaging from PEP and HerMES \citep{lutzPACS2011, oliverHerschel2012}, SCUBA-2 850 µm imaging from S2COSMOS \citep{caseyCharacterization2013, geachSCUBA22017, simpsonEast2019}, ALMA 1.2mm imaging from CHAMPS (over 0.18 deg$^2$; Faisst et al.~\textit{in prep}), ALMA 2 mm imaging from MORA \citep[over $\sim 0.2$ deg$^2$;][Long et al.~\textit{in prep}]{caseyMapping2021a}, and radio imaging from MeerKAT at 1.4 GHz \citep{heywoodMIGHTEE2022} and VLA at 3 GHz \citep{smolcicVLACOSMOS2017}.

\subsection{Catalog construction}\label{sec:catalog}

Details on the photometric catalogs for COSMOS-Web will be presented in a future paper (M. Shuntov, L.~Paquereau et al.~\textit{in prep.}); here, we provide a brief summary of the photometric catalog used in this work. 
We conduct source detection and aperture photometry using SEP \citep{barbarySEP2016}, the python implementation of SourceExtractor \citep{bertinSExtractor1996}.
We perform source detection on a $\chi^2$ detection image constructed from all four NIRCam bands \citep[e.g.][]{szalaySimultaneous1999}. 
First, F115W, F150W, and F277W are PSF-homogenized to F444W and converted to noise-equalized images by multiplying by the square root of the weight map. 
We then fit the negative tail of the pixel distribution to a Gaussian function to measure the per-pixel rms of the noise-equalized map. 
The maps are normalized by this rms and truncated at negative values to avoid false positives arising from multiple colocated negative noise peaks. 
The detection image is then computed by adding these PSF-homogenized, truncated signal-to-noise maps in quadrature.
The resulting $\sqrt{\chi^2}$ image can be directly related to the probability of an individual pixel value being drawn from the sky noise distribution \citep[see][]{szalaySimultaneous1999}.

We adopt a detection threshold of $\sqrt{\chi^2} = 3.698$ (with no background subtraction), equivalent to a 2.33$\sigma$ detection, or a 1\% probability of drawing a given pixel value from the noise distribution. 
We then perform aperture photometry on the available \HST/ACS, \JWST/NIRCam, and \JWST/MIRI imaging. 
Similar to \citet{finkelsteinCEERS2023}, we utilize small elliptical apertures  (with a Kron factor $k=1.1$) and correct to the total flux based on the default Kron aperture flux in F444W ($k=2.5$). 
ACS and NIRCam images are PSF-matched to F444W (as with the detection stage), and photometry is corrected for the F444W PSF flux falling outside of the large Kron aperture. 
For the MIRI bands, we adopt the same aperture sizes and apply a correction for the fractional flux loss due to the larger PSF. 
In particular, we compare the aperture flux in the native resolution F444W image to the flux after PSF-matching F444W to the MIRI F770W PSF, and apply this correction to the measured MIRI flux. 
%We construct separate catalogs for COSMOS-Web and PRIMER, and cross-match the two. 

Photometric uncertainties are adjusted using a random aperture method to account for correlated noise in the mosaics. 
For each band, and for a range of aperture sizes, we place 200,000 random circular apertures across the field, avoiding areas with detected sources, and measure the standard deviation of the measured fluxes by fitting the negative tail of the Gaussian distribution. 
Similar to \citet{finkelsteinCEERS2023} and \citet{riekeJADES2023}, we fit a power law of the form $\sigma_N = \alpha N^{\beta/2}$, where $N$ is the number of pixels in the aperture. 
All photometric uncertainties are then added in quadrature with this random aperture noise measurement $\sigma_N$ for the corresponding aperture size.

%
% 
%Herschel/PACS 100µm
%Herschel/PACS 160µm 
%Herschel/SPIRE 250µm 
%Herschel/SPIRE 350µm
%Herschel/SPIRE 500µm
%
%SCUBA2 450µm
%SCUBA2 850m
%ALMA archive? 
%
%VLA 1.4GHz
%MIGHTEE MEERKAT 1.28GHz
%VLA 3GHz

\section{Sample Selection \& Characterization}
\label{sec:sample}

From the COSMOS-Web data, we construct our sample of LRDs, fit them each to a series of SED models, and characterize the typical properties and volume density of the population. 
Before describing our sample selection and SED fitting methodology, we present a brief note of clarification on the motivation behind the strategy we adopt.
While, ideally, we would be able to fit each object with a composite galaxy/AGN SED model, decomposing the two components, the limited information available (largely due to the lack of spectroscopic redshifts) makes this kind of analysis prone to overfitting. 
Instead, we would like to limit the number of free parameters; 
rather than attempting to decompose the SEDs into galaxy/AGN components, we explore two extreme scenarios:
\begin{enumerate}[nosep,label=\emph{\alph*}),leftmargin=5mm]	
	\item \textit{QSO models:} the LRD population is primarily composed of dust-reddened AGN, with the optical continuum dominated by the direct thermal emission from the accretion disk ($\ll 1$ pc scales, temperatures $\gtrsim 10^5$ K) \citep[e.g.][]{labbeUNCOVER2023, mattheeLittle2024, greeneUNCOVER2024, kokorevCensus2024}. Note that in this scenario, the host galaxy may dominate in the rest-frame UV, but is negligible in the rest-frame optical. 
	\item \textit{Galaxy models:} the LRD population is primarily composed of compact/dusty starbursts, with the optical continuum dominated by the emission from young stellar populations \citep[e.g.][]{williamsGalaxies2023,perez-gonzalezWhat2024}. Note that this scenario does not preclude the presence of AGN in the population, but just requires that the red continuum emission in the optical is dominated by the stellar component. 
\end{enumerate}
These two scenarios may be thought of as ``edge cases''; while neither likely represents the true nature of the entire population, exploring the implications of either scenario yields valuable physical insight.  
In both cases, we assume that the characteristic blue UV slope of the LRDs is simply leakage or scattering through a clumpy obscuring medium, similar to what is seen in some lower-redshift DSFGs \citep[e.g.][]{caseyAre2014} and red quasars \citep[e.g.][]{glikmanHighly2023}, and we note that we do not select directly for this blue component.

\subsection{Sample selection}\label{sec:sample:selection}
 
Figure~\ref{fig:selection} outlines our color-compactness sample selection procedure. 
From our master photometric catalog described in \S\ref{sec:catalog}, we first downselect our catalog to objects which appear compact in F444W, defined via the compactness metric
\begin{equation}
	C_{444} = f_{444}(d=0\farcs2)/f_{444}(d=0\farcs5)
\end{equation}
We use F444W, despite the larger PSF relative to shorter-wavelength bands, since the LRDs are brightest in this band (and it is sometimes the only band with sufficient S/N to yield a reliable measure of compactness). 
Through a series of source injection simulations, we found that point sources reliably have $0.5<C_{444}<0.7$ provided the signal-to-noise is $\gtrsim 20$.
We select objects with $C_{444} > 0.5$, meaning that $\gtrsim 50\%$ of the F444W broadband emission within a $0\farcs5$ diameter aperture is contained within a $0\farcs2$ diameter aperture. 
We exclude objects with $C_{444} > 0.7$, which tend to be imaging artifacts (hot pixels), which don't follow the curve-of-growth associated with the PSF.
In Section~\ref{sec:sample:morpho} we fit these objects to 2D profiles to verify their compact/point-like nature.

\begin{figure}
	\centering
	\includegraphics[width=\linewidth]{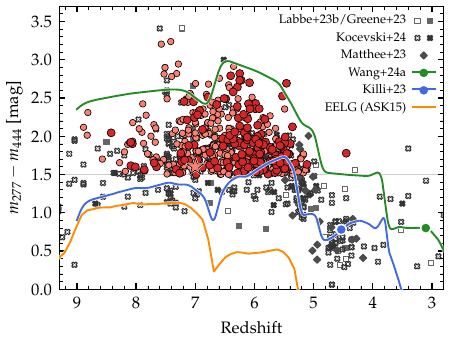}
	\caption{Comparison of our sample to literature LRD samples in F277W$-$F444W color vs.~redshift space. Our sample is shown in red, as in Figure~\ref{fig:selection}, and literature photometric/spectroscopic samples are shown in grey open/filled points, respectively \citep{labbeUNCOVER2023,greeneUNCOVER2024,mattheeLittle2024,kocevskiRise2024}. The blue and green points/lines highlight two lower-redshift LRDs with high-SNR PRISM spectra (J0647\_1045 at $z=4.5$ from \citealt{killiDeciphering2023} and RUBIES-BLAGN-1 at $z=3.1$ from \citealt{wangRUBIES2024}) and their F277W$-$F444W color when redshifted.
	The dark orange line shows a redshifted SDSS extreme emission line galaxy (EELG) template \citep[the ASK15 template from][]{sanchezalmeidaAutomatic2010}.
	We see that selecting for the reddest objects, ($m_{277}-m_{444}>1.5$, as indicated by the grey line) biases our sample to $z\gtrsim 5$ due to the shifting rest-frame wavelengths probed by F277W and F444W.
	The bimodal redshift distribution of our sample is a result of either \ha\ (from $z\sim 5$--$7$) or \oiii\,$\lambda 5007$ ($z\sim 7$--$9$) falling into F444W and contributing to the extremely red color, though the selection for $m_{277}-m_{444}>1.5$ mitigates contamination from intrinsically blue, strong emission line objects.}\label{fig:selection_comparison}
\end{figure}

We then select objects with red F277W--F444W colors, 
\begin{equation}
	m_{277}-m_{444} > 1.5.
\end{equation} 
This is the same threshold that was adopted by \citet{barroExtremely2024}, who found that almost all objects satisfying this color threshold appear point-like and blue from $1$--$2$\,\textmu m. 
Similarly, while \citet{mattheeLittle2024} selected for broad-line AGN based on the H$\alpha$ emission, the majority of their objects show similarly red colors from $2$ to $\sim 3$--$5$ \textmu m. 
We note that this color threshold is more stringent than several recent works which select LRDs via F277W$-$F444W $> 0.7$ or F200W$-$F356W $>1.0$ \citep[e.g.][]{labbeUNCOVER2023,kokorevCensus2024}.
We focus on the reddest subset of the LRD population, which may bias our sample towards more dust-obscured and/or higher-redshift objects. 
Figure~\ref{fig:selection_comparison} shows our sample compared to other literature samples in color-redshift space; here, we adopt the redshifts from QSO SED models as described in section~\ref{sec:sedfitting:qso}. 
We also show in Figure~\ref{fig:selection_comparison} the F277W$-$F444W color vs.~redshift derived by redshifting the high SNR NIRSpec/PRISM spectra of two lower-redshift LRDs (J0647\_1045 at $z=4.532$ from \citet{killiDeciphering2023} and RUBIES-BLAGN-1 at $z=3.104$ from \citet{wangRUBIES2024}). 
Indeed, by selecting for the reddest objects, we bias our sample to $z\gtrsim 5$, where F277W$-$F444W covers the rest-frame optical range. 
Moreover, the somewhat bimodal redshift distribution is a result of either \ha\ (from $z\sim 5$--$7$) or \oiii\,$\lambda 5007$ ($z\sim 7$--$9$) falling into F444W and contributing to the extremely red color. 

We note that our selection is perhaps better described as a selection for extremely red objects \citep[EROs, as in][]{barroExtremely2024}, as LRDs are often selected based on the composite blue+red spectral shape \citep[e.g.][]{labbeUNCOVER2023,kokorevCensus2024,kocevskiRise2024}. 
However, as the physical origin of the blue UV component is unclear, requiring it in selection may bias the sample towards objects with unobscured host galaxies or unique dust geometries. 
Because of this, and due to the shallow depth of the F115W+F150W imaging in COSMOS-Web ($5\sigma \sim 27.1$--$27.6$ AB mag), we opt to use only the red F277W--F444W criterion. 
We also note that other LRD studies \citep[e.g.][]{labbeUNCOVER2023,kokorevCensus2024} often require red colors in multiple filter pairs to rule out contamination from extreme emission line galaxies (EELGs) which may exhibit a very red F277W--F444W color despite a blue continuum.
While our selection has the potential to yield such contaminants, we note that by focusing on the reddest objects ($\sim 0.5$ mag redder than other selections), we largely mitigate this effect; in fact, $\sim 90\%$ of our sample that is covered by the PRIMER survey ($N=37$) satisfy the multi-band color selection described in \citet{labbeUNCOVER2023}. 
If we instead adopted a less stringent single color criteria, F277W$-$F444W$>0.7$, this fraction drops to $65\%$.

The simple selection criteria described above yields a sample of 533 sources across the 0.54 sq.~degree COSMOS-Web field. 
Of these, 163 sources are covered with MIRI F770W imaging (32 of which also have F1800W imaging from PRIMER), and 39 are covered with PRIMER NIRCam imaging.

\subsection{Removal of Likely Brown Dwarfs}

In order to remove contaminanting brown dwarf stars in the Milky Way \citep[e.g.][]{langeroodiLittle2023,hainlineBrown2024,burgasserUNCOVER2024}, we fit each candidate with a grid of low-temperature stellar atmosphere models.
We construct a grid of models that spans a large parameter space, including cloudy, chemical equilibrium models \citep[Sonora-Diamondback, $T\sim 900$--$2400$ K;][]{morleySonora2024}, cloud-free models in both chemical equilibrium (Sonora-Bobcat, $T\sim 200$--$1300$ K; \citealt{marleySonora2021}) and disequilibrium (ElfOwl, $T\sim 300$--$1000$ K; \citealt{mukherjeeSonora2024}), and low-metallicity models (LOWZ, $[{\rm M}/{\rm H}]=-1$; \citealt{meisnerNew2021}). 
We explore a range of temperatures $T\sim 200$--$2400$ K, surface gravities $g=100$ and $3160$, and metallicities $[{\rm M}/{\rm H}] = -1$, $-0.5$, and $0$. 
We convolve the model SEDs with the \textit{HST}+\textit{JWST} filter curves and perform a simple grid-fitting routine, scaling the fluxes of each model to minimize the $\chi^2$. 
We adopt the model with the lowest $\chi^2$ as the best-fitting stellar model, and compare with the best-fit galaxy/quasar SEDs described later in Section~\ref{sec:sedfitting}. 
104 objects in our sample have brown dwarf model $\chi^2$ less than the minimum galaxy/quasar model $\chi^2$ and are removed. 
The fraction of sources estimated to be brown dwarfs is $\sim 9\%$ for the MIRI subsample, and $\sim 24\%$ for the NIRCam-only sample.
This difference is due to the presence of strong absorption bands in cold brown dwarfs at $\sim 8$\,\textmu m, allowing MIRI/F770W imaging to break degeneracies between galaxy/QSO and brown dwarf models, particularly that our NIRCam data have only four bands.

After the brown dwarf removal procedure, our final LRD sample contains 434 sources, 148 of which are covered with MIRI F770W imaging and 29 of which are covered with PRIMER NIRCam imaging.

\subsection{Morphological Fitting}\label{sec:sample:morpho}

\begin{figure}
\centering
\includegraphics[width=\linewidth]{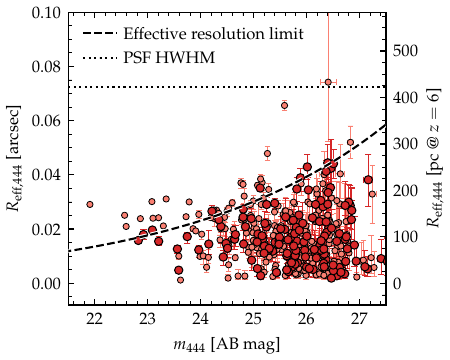}
\caption{Effective radius vs.~F444W magnitude for our sample. The dotted line indicates the half-width at half-maximum (HWHM) for the F444W PSF. The dashed line indicates the effective resolution limit based on the measured sizes of PSF stars. While the majority of our sample is unresolved, a few sources appear marginally resolved, perhaps indicating host galaxy contribution.}\label{fig:reff}	
\end{figure}

We perform 2D image fitting to the \JWST/NIRCam imaging to validate the simplified ``compactness'' selection and determine whether any sources are marginally resolved. 
We do this using \textsc{galsim}, a galaxy image forward-modeling code designed to produce realistic simulated images based on analytical profiles \citep{roweGalSim2015}. 
We opt for this approach, rather than the commonly-used image fitting tools (such as \textsc{galfit}; \citealt{pengDetailed2002, pengDetailed2010}), in order to conduct Bayesian inference; specifically, we utilize the nested sampling algorithim \textsc{MultiNest} \citep{ferozMultimodal2008,ferozMULTINEST2009,ferozImportance2019} to fully explore the parameter space and obtain robust uncertainties on the derived sizes for each object.
We fit the F444W imaging for each object to a PSF-convolved 2D Sérsic profile with the effective radius allowed to vary from $0\farcs001$ to $0\farcs3$, the Sérsic index $n$ from $0.5$ to $4$, and the axis ratio $q$ from $0.7$ to $1$. 
Figure~\ref{fig:reff} shows the measured F444W effective radius vs.~the F444W magnitude for our sample. 
The half-width at half-maximum for the F444W PSF is shown in the black dotted line, as an approximate resolution limit for the observations. 
We can constrain the sizes even below the PSF size, for high SNR sources, as even slightly extended objects would show some residuals after PSF subtraction. 
To quantify this effect, we performed our Sérsic profile fitting to PSF stars identified via their position in the half-light radius vs.~magnitude plane. 
To avoid including marginally resolved sources in this selection, at $m_{444}>24$, we adopt the brown dwarf sample (identified as contaminants in Section~\ref{sec:sample:selection}) to estimate the limiting size. 
We then fit a curve to the 90th percentile of the sizes measured in bins of F444W magnitude, deriving effective resolution limit shown in the dashed line in Figure~\ref{fig:reff}.

Nearly all of our candidates are unresolved, which is to be expected, as we are selecting for compact sources based on the F444W aperture photometry. 
The unresolved nature of the LRDs places an upper limit on their physical sizes, which ranges from $R_{\rm eff} \lesssim 0\farcs01$ at $m_{444} \approx 22$ to $\lesssim 0\farcs04$ at $m_{444} \approx 27$. 
At $z=6$, this corresponds to effective radii $\lesssim 100$--$300$ pc. 
A few sources are marginally resolved, particularly the brightest subset ($m_{\rm 444} \lesssim 24$), perhaps indicating some host galaxy contribution. 
We keep these marginally resolved sources in the sample, as they satisfy the original compactness criteria and have effective radii $\lesssim 200$ pc. 

%
%However, a few objects are marginally resolved; we highlight a handful of these in the RGB plots in the right panels. 
%While the unresolved sources (yellow) show clear PSF features, the marginally resolved sources (green) are less clear. 
%A number of sources are fit to marginally resolved models due to the presence of a close companion (blue). 
%Given their close proximity and similar colors, it is plausible that these companions are physically associated with the EROs in question, though it is not certain. 
%

\subsection{Spectral Energy Distribution Fitting}\label{sec:sedfitting}

We fit both galaxy and QSO models to all objects in our sample. 
However, we do not intend to use these results to classify objects as either galaxy or AGN-dominated based on goodness-of-fit---there are, in general more free parameters in the model than we have data points.
Instead, as we are testing the validity of either extreme scenario (the ``edge cases''), intend to examine the results from the model fitting under the assumption of either scenario.

\begin{figure*}
 \centering
 \epsscale{1.15}
 \plottwo{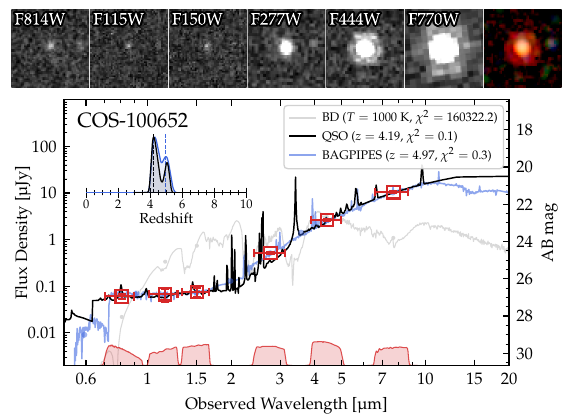}{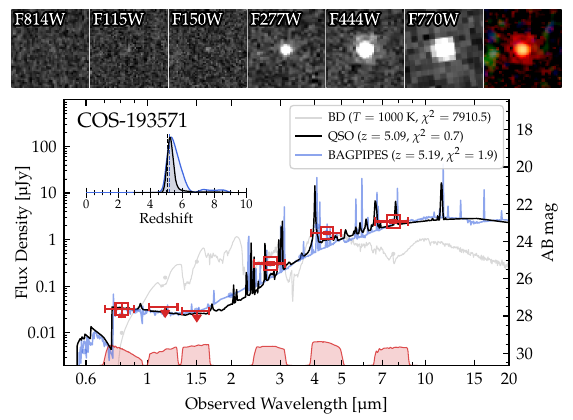}
 \plottwo{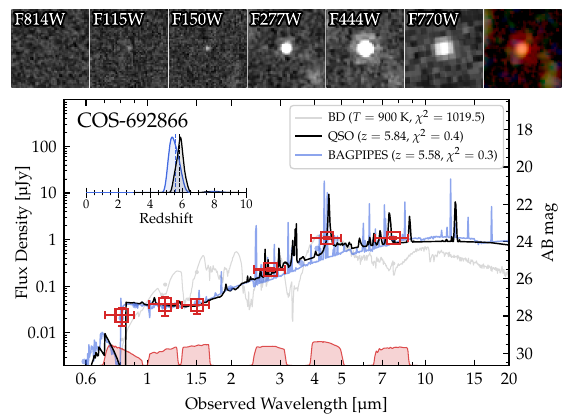} {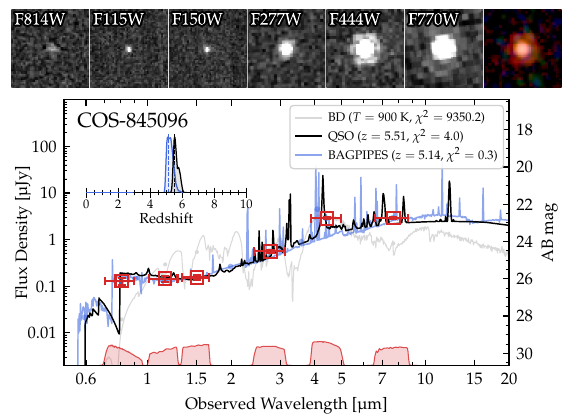}
 \plottwo{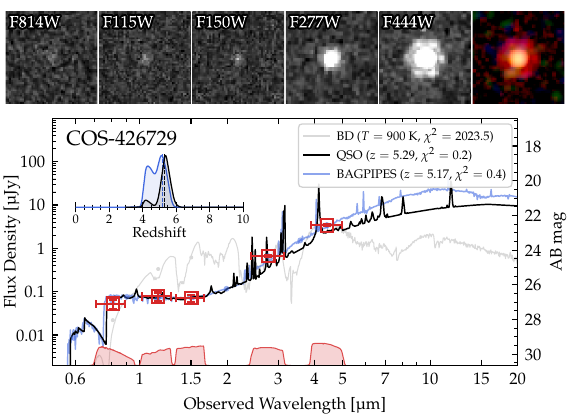}{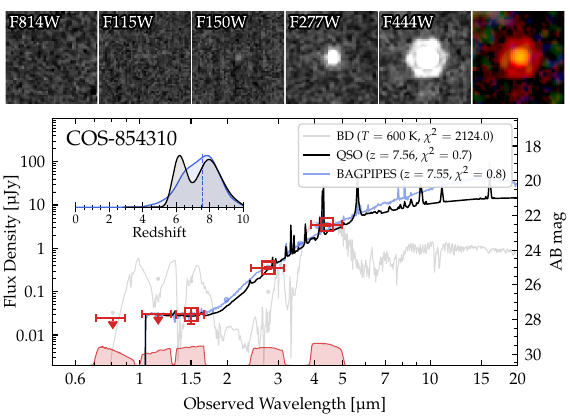}
 \caption{Cutouts and SEDs for six particularly bright ($m_{444} \lesssim 24$) LRDs. We show 1.5'' square cutouts in \HST/ACS F814W, \JWST/NIRCam F115W, F150W, F277W, and F444W, and \JWST/MIRI F770W (where available). 
Measured photometry is shown in red, or 2$\sigma$ upper limits for non-detections; the best-fitting QSO model is shown in black, \texttt{bagpipes} in blue, and brown dwarf in light grey. 
The inset panels show the redshift probability distribution with the maximum likelihood value marked with a dashed line. 
The complete figure set (434 images) is available at \url{https://github.com/hollisakins/akins24_cw}.}\label{fig:SEDs1}
\end{figure*} 
% The complete figure set (434 images) is available in the online journal.

\begin{figure*}[t!]
 \centering
 \epsscale{1.15}
 \plottwo{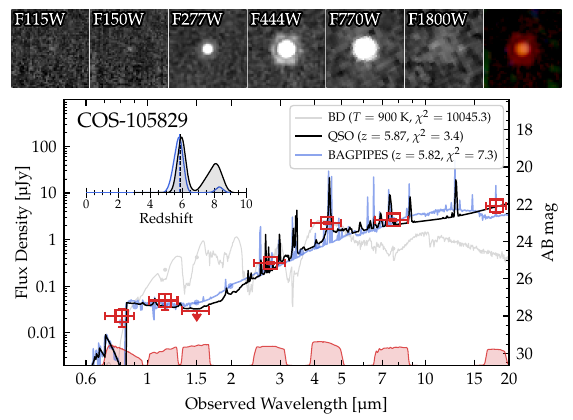}{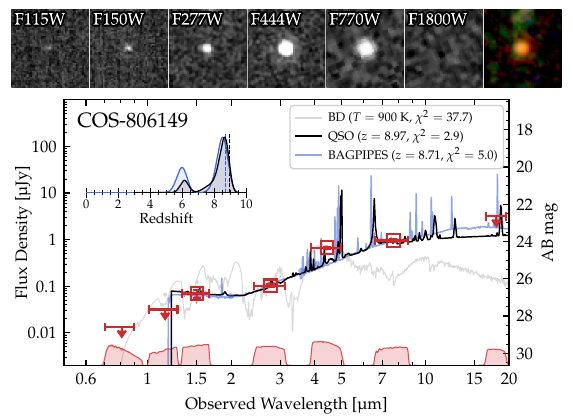}
  \plottwo{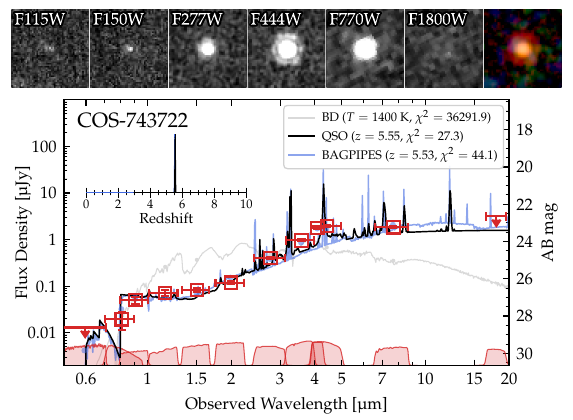}{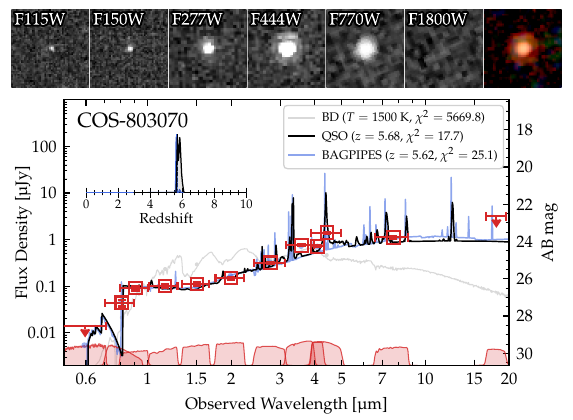}
 \caption{Same as Figure~\ref{fig:SEDs1}, but for four bright LRDs with MIRI F1800W coverage from PRIMER. Two sources (above) are only in COSMOS-Web NIRCam coverage, but fall within the PRIMER MIRI coverage; two sources (below) have both PRIMER NIRCam and MIRI coverage. In all cases, the MIRI F1800W imaging yields marginal or non-detections, consistent with a flat SED in $f_\nu$ from rest-frame $\sim 1$--$3$\,\textmu m.}\label{fig:SEDs2}
\end{figure*}

\subsubsection{QSO models}\label{sec:sedfitting:qso}

We first fit each LRD in the sample to Type I AGN models. 
In particular, we adopt broad-line AGN templates, consistent with the high fraction of broad-line AGN confirmed in LRDs observed with \JWST\ spectroscopy \citep[e.g.][]{mattheeLittle2024, greeneUNCOVER2024}. 
While existing SED fitting codes such as CIGALE \citep{burgarellaStar2005, nollAnalysis2009, boquienCIGALE2019} or \textsc{prospector} \citep{lejaDeriving2017a, johnsonStellar2021} do include AGN models, these are primarily designed to decompose host galaxy/AGN components, particularly in the rest-frame mid-IR, and thus are not ideal for this study. 
Moreover, these codes do not typically include strong/broad emission lines from the AGN, which may contribute significantly to the observed photometry. 
We therefore employ a custom SED fitting code, assuming that the emission is dominated by AGN, and employing flexibile dust attenuation and emission line models.

We model the intrinsic accretion disk emission as a single power law with a slope $\beta$ ($f_\lambda \propto \lambda^\beta$).  
The normalization is set by the intrinsic UV magnitude $M_{\rm UV,int}$, prior to any dust attenuation. We allow $M_{\rm UV,int}$ to vary from $-18$ to $-27$ and $\beta$ to vary from $-1.5$ to $-2.8$, both with flat priors.

We employ the \citet{templeModelling2021} QSO emission line template with \verb|emline_type = 0|, corresponding to the average emission line template at $z=2$, though we employ a few adjustments. 
In particular, we allow the overall emission line scaling to vary from $1$ to $2$ times the nominal value, which is calibrated to a reference QSO continuum. 
This is motivated by the higher broad-line EWs observed for \JWST-selected AGN compared to local samples \citep[see e.g.][]{maiolinoJWST2024}. 
We also disable Lyman-$\alpha$ emission due to the opacity of the IGM at the redshifts relevant to our study (though we note that several LRDs exhibit Ly$\alpha$ emission, even at $z>7$, e.g. \citealt{furtakHigh2024,kokorevUNCOVER2023}).
Even at $z<7$, this choice shouldn't impact the results significantly, given the relatively low EWs typical of Ly$\alpha$ emission \citep[$\lesssim 100$\,\AA,][]{kobayashiLya2010} compared to the filter bandwidths ($\sim 3000$\,\AA).

We additionally use a flexible dust attenuation prescription to account for the blue UV slope characteristic of this population. 
We adopt the QSO attenuation law from \citet{templeModelling2021}, which is similar to the SMC extinction curve (except at $\lambda \lesssim 1700$ \AA, where it is significantly shallower). 
We add an additional parameter, a ``scattering'' fraction, or a fraction of the intrinsic spectrum that escapes unattenuated. We allow this scattering fraction to vary from $10^{-5}$ to $10^{-1}$ with a log-uniform prior. 
The effect of this is similar to employing an even greyer attenuation law, but adds a bit more

\begin{longrotatetable}
%  \startlongtable
%\begin{deluxetable*}{@{\extracolsep{8pt}}l@{}C@{}C@{}C@{}C@{}C@{}D@{}D@{}D@{}D@{}D@{}D@{}}
\begin{deluxetable*}{lCCCCCDDDDDDD}
\tablecaption{Photometric measurements and derived physical properties for the COSMOS-Web LRDs}\label{tab:sources}
\tablewidth{670pt}
\tabletypesize{\scriptsize}
%\colnumbers
\tablehead{\colhead{\multirow{2}{*}{ID}} & \colhead{R.A.} & \colhead{Decl.} & \colhead{F277W} & \colhead{F444W} & \colhead{F770W} & \multicolumn{6}{c}{QSO fitting} & \multicolumn{6}{c}{BAGPIPES} & \twocolhead{$R_{\rm eff,444}$} \\[-4pt]
	\cmidrule(lr){7-12}\cmidrule(lr){13-18}\\[-18pt]
~ & \text{[J2000]} & \text{[J2000]} & \text{[AB mag]} & \text{[AB mag]} & \text{[AB mag]} & \twocolhead{$z_{\rm qso}$} & \twocolhead{$\log L_{\rm bol}$} & \twocolhead{$A_V$} & \twocolhead{$z_{\rm gal}$} & \twocolhead{$\log M_\star$} & \twocolhead{$A_V$} & \twocolhead{[mas]} }
\decimals
\startdata
1491 & 09^{\rm h}59^{\rm m}22_{.}^{\rm s}58 & +02^{\rm d}07^{\rm m}43_{.}^{\rm s}17 & 26.45 & 24.77 & \nodata & $6.2_{-0.5}^{+1.2}$ & $46.2_{-0.3}^{+0.4}$ & $3.4_{-0.5}^{+0.7}$ & $6.1_{-0.7}^{+1.2}$ & $10.7_{-0.3}^{+0.3}$ & $3.3_{-0.7}^{+0.8}$ & $19.3^{+2.8}_{-3.6}$ \\
3962 & 09^{\rm h}59^{\rm m}08_{.}^{\rm s}80 & +02^{\rm d}10^{\rm m}48_{.}^{\rm s}23 & 26.24 & 24.66 & \nodata & $7.3_{-1.8}^{+1.9}$ & $46.2_{-0.4}^{+0.3}$ & $2.2_{-0.3}^{+0.6}$ & $7.0_{-2.3}^{+2.1}$ & $10.4_{-0.2}^{+0.3}$ & $2.0_{-0.5}^{+1.0}$ & $37.1^{+1.5}_{-2.1}$ \\
4056 & 09^{\rm h}59^{\rm m}27_{.}^{\rm s}39 & +02^{\rm d}09^{\rm m}10_{.}^{\rm s}80 & 27.15 & 25.37 & 24.98 & $6.0_{-0.5}^{+1.0}$ & $45.8_{-0.2}^{+0.2}$ & $2.9_{-0.5}^{+0.5}$ & $6.2_{-0.7}^{+2.2}$ & $9.4_{-0.3}^{+0.3}$ & $2.2_{-0.5}^{+0.5}$ & $4.0^{+3.6}_{-2.0}$ \\
8203 & 09^{\rm h}59^{\rm m}18_{.}^{\rm s}51 & +02^{\rm d}13^{\rm m}25_{.}^{\rm s}82 & 26.96 & 25.20 & 26.07 & $8.3_{-0.4}^{+0.4}$ & $46.2_{-0.3}^{+0.2}$ & $5.1_{-1.9}^{+0.6}$ & $5.3_{-0.2}^{+0.2}$ & $8.7_{-0.1}^{+0.1}$ & $1.7_{-0.2}^{+0.3}$ & $28.4^{+1.8}_{-2.4}$ \\
9430 & 09^{\rm h}59^{\rm m}13_{.}^{\rm s}39 & +02^{\rm d}14^{\rm m}38_{.}^{\rm s}54 & 27.44 & 25.84 & 25.24 & $7.3_{-1.2}^{+1.0}$ & $45.9_{-0.3}^{+0.3}$ & $3.1_{-0.5}^{+0.8}$ & $7.8_{-0.6}^{+0.6}$ & $9.8_{-0.3}^{+0.2}$ & $1.9_{-0.4}^{+0.5}$ & $15.7^{+4.2}_{-5.0}$ \\
9680 & 09^{\rm h}59^{\rm m}18_{.}^{\rm s}44 & +02^{\rm d}14^{\rm m}20_{.}^{\rm s}41 & 27.51 & 25.27 & 24.88 & $7.5_{-1.4}^{+0.7}$ & $46.3_{-0.3}^{+0.3}$ & $3.8_{-0.6}^{+1.4}$ & $5.8_{-0.4}^{+1.6}$ & $9.3_{-0.2}^{+0.2}$ & $3.0_{-0.6}^{+0.4}$ & $11.9^{+4.2}_{-3.8}$ \\
10023 & 09^{\rm h}59^{\rm m}11_{.}^{\rm s}55 & +02^{\rm d}15^{\rm m}09_{.}^{\rm s}29 & 27.69 & 26.04 & 26.01 & $6.0_{-0.4}^{+1.0}$ & $45.3_{-0.2}^{+0.2}$ & $2.4_{-0.4}^{+0.5}$ & $6.0_{-0.6}^{+2.3}$ & $8.9_{-0.2}^{+0.3}$ & $1.9_{-0.5}^{+0.5}$ & $34.4^{+4.5}_{-4.9}$ \\
10511 & 09^{\rm h}59^{\rm m}17_{.}^{\rm s}57 & +02^{\rm d}14^{\rm m}53_{.}^{\rm s}62 & 26.53 & 25.03 & 25.30 & $5.9_{-0.3}^{+0.4}$ & $45.6_{-0.1}^{+0.1}$ & $1.9_{-0.3}^{+0.3}$ & $5.5_{-0.3}^{+0.6}$ & $8.9_{-0.1}^{+0.2}$ & $1.7_{-0.3}^{+0.2}$ & $18.2^{+1.7}_{-1.4}$ \\
11737 & 09^{\rm h}59^{\rm m}26_{.}^{\rm s}60 & +02^{\rm d}14^{\rm m}49_{.}^{\rm s}10 & 27.38 & 25.23 & \nodata & $7.8_{-1.8}^{+1.5}$ & $46.5_{-0.4}^{+0.5}$ & $3.6_{-0.8}^{+1.1}$ & $7.5_{-2.3}^{+1.7}$ & $10.7_{-0.4}^{+0.7}$ & $3.3_{-0.8}^{+1.1}$ & $29.1^{+1.7}_{-1.8}$ \\
13618 & 09^{\rm h}59^{\rm m}18_{.}^{\rm s}95 & +02^{\rm d}16^{\rm m}51_{.}^{\rm s}10 & 28.10 & 26.34 & 25.91 & $6.0_{-0.5}^{+0.6}$ & $45.4_{-0.2}^{+0.2}$ & $2.8_{-0.5}^{+0.6}$ & $6.2_{-0.8}^{+2.1}$ & $9.1_{-0.3}^{+0.5}$ & $2.1_{-0.6}^{+0.4}$ & $28.2^{+5.7}_{-6.9}$ \\
14599 & 09^{\rm h}59^{\rm m}23_{.}^{\rm s}86 & +02^{\rm d}16^{\rm m}57_{.}^{\rm s}03 & 27.16 & 25.55 & 25.59 & $5.9_{-0.4}^{+0.4}$ & $45.5_{-0.2}^{+0.2}$ & $2.3_{-0.4}^{+0.4}$ & $5.6_{-0.4}^{+2.1}$ & $8.9_{-0.2}^{+0.3}$ & $1.9_{-0.4}^{+0.3}$ & $19.9^{+3.0}_{-3.7}$ \\
15495 & 09^{\rm h}59^{\rm m}39_{.}^{\rm s}19 & +02^{\rm d}16^{\rm m}04_{.}^{\rm s}16 & 27.39 & 25.18 & 25.20 & $8.1_{-0.7}^{+0.5}$ & $45.9_{-0.2}^{+0.2}$ & $2.4_{-0.3}^{+0.7}$ & $5.5_{-0.3}^{+0.5}$ & $9.1_{-0.1}^{+0.2}$ & $2.6_{-0.2}^{+0.2}$ & $16.5^{+4.2}_{-4.7}$ \\
15654 & 09^{\rm h}59^{\rm m}39_{.}^{\rm s}29 & +02^{\rm d}16^{\rm m}08_{.}^{\rm s}81 & 26.48 & 24.72 & 24.11 & $5.7_{-0.3}^{+0.4}$ & $46.2_{-0.2}^{+0.2}$ & $3.6_{-0.5}^{+0.5}$ & $5.6_{-0.3}^{+0.6}$ & $9.8_{-0.2}^{+0.2}$ & $2.8_{-0.3}^{+0.3}$ & $24.4^{+1.9}_{-1.8}$ \\
15961 & 09^{\rm h}59^{\rm m}15_{.}^{\rm s}84 & +02^{\rm d}18^{\rm m}26_{.}^{\rm s}22 & 26.71 & 25.08 & \nodata & $6.2_{-0.5}^{+0.9}$ & $46.3_{-0.3}^{+0.4}$ & $4.2_{-0.9}^{+1.1}$ & $6.4_{-0.6}^{+1.2}$ & $10.9_{-0.4}^{+0.4}$ & $4.2_{-1.0}^{+1.1}$ & $17.2^{+3.4}_{-4.2}$ \\
16108 & 09^{\rm h}59^{\rm m}17_{.}^{\rm s}01 & +02^{\rm d}18^{\rm m}26_{.}^{\rm s}10 & 27.99 & 25.77 & 25.25 & $6.0_{-0.4}^{+1.3}$ & $45.8_{-0.2}^{+0.2}$ & $3.6_{-0.6}^{+0.6}$ & $5.6_{-0.4}^{+0.5}$ & $9.2_{-0.2}^{+0.2}$ & $3.2_{-0.5}^{+0.4}$ & $7.6^{+5.5}_{-4.4}$ \\
16692 & 09^{\rm h}59^{\rm m}04_{.}^{\rm s}68 & +02^{\rm d}08^{\rm m}20_{.}^{\rm s}30 & 28.43 & 26.75 & \nodata & $6.6_{-1.3}^{+2.0}$ & $45.5_{-0.4}^{+0.4}$ & $3.1_{-0.9}^{+1.7}$ & $6.5_{-1.5}^{+2.1}$ & $9.7_{-0.4}^{+0.5}$ & $2.8_{-1.0}^{+1.6}$ & $28.5^{+8.5}_{-9.7}$ \\
17632 & 09^{\rm h}59^{\rm m}04_{.}^{\rm s}72 & +02^{\rm d}08^{\rm m}56_{.}^{\rm s}02 & 29.97 & 26.73 & \nodata & $7.7_{-1.5}^{+1.1}$ & $46.0_{-0.4}^{+0.4}$ & $4.3_{-1.0}^{+0.9}$ & $6.9_{-1.5}^{+1.8}$ & $10.3_{-0.5}^{+0.6}$ & $4.2_{-1.2}^{+1.1}$ & $25.1^{+6.6}_{-8.6}$ \\
18065 & 09^{\rm h}59^{\rm m}05_{.}^{\rm s}10 & +02^{\rm d}09^{\rm m}07_{.}^{\rm s}29 & 27.54 & 25.95 & \nodata & $5.9_{-0.5}^{+1.9}$ & $45.8_{-0.4}^{+0.3}$ & $3.9_{-1.4}^{+1.3}$ & $5.2_{-0.7}^{+1.9}$ & $10.0_{-0.4}^{+0.4}$ & $4.1_{-1.5}^{+1.2}$ & $26.9^{+4.4}_{-4.0}$ \\
19068 & 09^{\rm h}59^{\rm m}07_{.}^{\rm s}68 & +02^{\rm d}09^{\rm m}24_{.}^{\rm s}36 & 28.69 & 27.18 & \nodata & $6.5_{-1.4}^{+2.1}$ & $45.2_{-0.4}^{+0.3}$ & $2.8_{-0.8}^{+1.9}$ & $6.7_{-1.6}^{+2.0}$ & $9.4_{-0.3}^{+0.4}$ & $2.2_{-0.7}^{+1.2}$ & $9.5^{+8.2}_{-5.7}$ \\
26156 & 09^{\rm h}59^{\rm m}32_{.}^{\rm s}88 & +02^{\rm d}11^{\rm m}00_{.}^{\rm s}35 & 28.39 & 26.28 & \nodata & $7.3_{-1.5}^{+1.8}$ & $46.0_{-0.5}^{+0.6}$ & $3.8_{-1.0}^{+1.3}$ & $7.0_{-1.8}^{+1.9}$ & $10.3_{-0.5}^{+0.7}$ & $3.7_{-1.1}^{+1.2}$ & $27.0^{+5.7}_{-6.5}$ \\
27062 & 09^{\rm h}59^{\rm m}17_{.}^{\rm s}69 & +02^{\rm d}12^{\rm m}52_{.}^{\rm s}84 & 28.82 & 27.28 & \nodata & $6.1_{-0.9}^{+1.8}$ & $45.3_{-0.4}^{+0.4}$ & $3.7_{-1.3}^{+1.6}$ & $5.7_{-1.1}^{+2.2}$ & $9.4_{-0.4}^{+0.5}$ & $3.3_{-1.4}^{+1.7}$ & $5.8^{+5.9}_{-3.1}$ \\
27100 & 09^{\rm h}59^{\rm m}14_{.}^{\rm s}24 & +02^{\rm d}13^{\rm m}13_{.}^{\rm s}22 & 28.59 & 26.42 & \nodata & $6.9_{-1.2}^{+1.7}$ & $45.8_{-0.4}^{+0.4}$ & $3.7_{-1.0}^{+1.3}$ & $6.7_{-1.6}^{+1.7}$ & $10.2_{-0.5}^{+0.6}$ & $3.5_{-1.0}^{+1.2}$ & $23.3^{+6.6}_{-6.5}$ \\
29753 & 09^{\rm h}59^{\rm m}19_{.}^{\rm s}49 & +02^{\rm d}14^{\rm m}10_{.}^{\rm s}35 & 28.35 & 26.48 & 25.68 & $6.0_{-0.6}^{+1.6}$ & $45.6_{-0.3}^{+0.2}$ & $3.6_{-1.0}^{+0.9}$ & $6.8_{-1.3}^{+1.5}$ & $9.6_{-0.3}^{+0.3}$ & $2.1_{-0.5}^{+0.7}$ & $2.8^{+1.8}_{-1.4}$ \\
36933 & 09^{\rm h}59^{\rm m}19_{.}^{\rm s}29 & +02^{\rm d}17^{\rm m}12_{.}^{\rm s}99 & 28.59 & 26.97 & 26.32 & $6.4_{-0.9}^{+1.6}$ & $45.3_{-0.3}^{+0.3}$ & $3.1_{-0.9}^{+1.2}$ & $7.2_{-1.5}^{+1.3}$ & $9.3_{-0.3}^{+0.2}$ & $1.8_{-0.4}^{+0.6}$ & $28.7^{+7.2}_{-8.3}$ \\
37490 & 09^{\rm h}59^{\rm m}30_{.}^{\rm s}92 & +02^{\rm d}16^{\rm m}24_{.}^{\rm s}49 & 28.52 & 26.86 & 25.83 & $6.5_{-1.2}^{+1.8}$ & $45.6_{-0.3}^{+0.3}$ & $3.6_{-1.2}^{+1.6}$ & $6.4_{-1.4}^{+1.9}$ & $9.5_{-0.3}^{+0.3}$ & $2.2_{-0.6}^{+0.8}$ & $29.9^{+6.6}_{-8.2}$ \\
39192 & 09^{\rm h}59^{\rm m}13_{.}^{\rm s}42 & +02^{\rm d}18^{\rm m}47_{.}^{\rm s}20 & 28.52 & 26.65 & \nodata & $6.7_{-1.1}^{+2.0}$ & $45.7_{-0.4}^{+0.4}$ & $3.3_{-0.8}^{+1.5}$ & $6.4_{-1.5}^{+2.0}$ & $9.9_{-0.4}^{+0.6}$ & $3.3_{-1.1}^{+1.4}$ & $74.2^{+25.6}_{-22.0}$ \\
42832 & 09^{\rm h}59^{\rm m}50_{.}^{\rm s}78 & +02^{\rm d}05^{\rm m}43_{.}^{\rm s}96 & 28.41 & 26.41 & \nodata & $7.6_{-1.5}^{+1.7}$ & $46.0_{-0.4}^{+0.5}$ & $3.7_{-0.9}^{+1.2}$ & $7.9_{-2.2}^{+1.4}$ & $10.4_{-0.6}^{+1.0}$ & $3.6_{-1.1}^{+1.2}$ & $3.3^{+3.3}_{-1.6}$ \\
42963 & 10^{\rm h}00^{\rm m}01_{.}^{\rm s}24 & +02^{\rm d}04^{\rm m}50_{.}^{\rm s}94 & 28.16 & 26.27 & \nodata & $8.1_{-1.9}^{+1.3}$ & $46.1_{-0.5}^{+0.6}$ & $3.7_{-0.9}^{+1.3}$ & $8.1_{-2.3}^{+1.0}$ & $10.5_{-0.6}^{+0.9}$ & $3.3_{-1.0}^{+1.4}$ & $3.4^{+3.4}_{-1.6}$ \\
44171 & 09^{\rm h}59^{\rm m}37_{.}^{\rm s}77 & +02^{\rm d}07^{\rm m}38_{.}^{\rm s}65 & 27.47 & 25.75 & \nodata & $7.2_{-1.6}^{+2.0}$ & $45.9_{-0.4}^{+0.4}$ & $2.6_{-0.5}^{+0.9}$ & $6.9_{-2.0}^{+1.9}$ & $10.1_{-0.3}^{+0.4}$ & $2.4_{-0.6}^{+1.0}$ & $8.3^{+6.3}_{-4.5}$ \\
46758 & 09^{\rm h}59^{\rm m}54_{.}^{\rm s}88 & +02^{\rm d}07^{\rm m}34_{.}^{\rm s}62 & 25.16 & 23.51 & 23.62 & $5.7_{-0.2}^{+0.2}$ & $46.3_{-0.1}^{+0.1}$ & $2.5_{-0.2}^{+0.2}$ & $5.2_{-0.2}^{+0.2}$ & $9.6_{-0.1}^{+0.1}$ & $2.0_{-0.1}^{+0.1}$ & $4.8^{+0.4}_{-0.0}$ \\
47467 & 09^{\rm h}59^{\rm m}41_{.}^{\rm s}90 & +02^{\rm d}09^{\rm m}07_{.}^{\rm s}89 & 26.53 & 24.92 & 24.12 & $5.4_{-0.2}^{+0.3}$ & $46.2_{-0.2}^{+0.2}$ & $4.3_{-0.6}^{+0.6}$ & $5.4_{-0.3}^{+0.4}$ & $9.9_{-0.2}^{+0.2}$ & $3.0_{-0.4}^{+0.4}$ & $23.9^{+3.6}_{-3.8}$ \\
47745 & 09^{\rm h}59^{\rm m}58_{.}^{\rm s}60 & +02^{\rm d}07^{\rm m}46_{.}^{\rm s}94 & 27.27 & 25.70 & 25.48 & $5.9_{-0.4}^{+0.5}$ & $45.5_{-0.2}^{+0.2}$ & $2.7_{-0.5}^{+0.5}$ & $5.8_{-0.5}^{+2.3}$ & $9.1_{-0.3}^{+0.4}$ & $2.2_{-0.7}^{+0.5}$ & $38.9^{+3.4}_{-5.1}$ \\
48256 & 10^{\rm h}00^{\rm m}03_{.}^{\rm s}90 & +02^{\rm d}07^{\rm m}34_{.}^{\rm s}23 & 27.14 & 24.71 & 24.17 & $6.3_{-0.4}^{+1.8}$ & $46.3_{-0.2}^{+0.1}$ & $3.5_{-0.7}^{+0.5}$ & $5.7_{-0.4}^{+0.7}$ & $9.6_{-0.2}^{+0.2}$ & $3.2_{-0.4}^{+0.3}$ & $14.5^{+5.1}_{-3.1}$ \\
\enddata
\tablecomments{Table~\ref{tab:sources} is available in its entirety in machine-readable format at \url{https://github.com/hollisakins/akins24_cw}. A portion is shown here for guidance regarding its form and content.}
\end{deluxetable*}
\end{longrotatetable}

\noindent flexibility and aids in the interpretation of the blue slope as coming from scattering around the AGN torus \citep[e.g.][]{labbeUNCOVER2023} or leakage through holes in the obscuring medium, perhaps created by outflows \citep[e.g.][]{noboriguchiSimilarity2023}. 
We additionally include a parameter, $\eta_{\rm NLR}$, representing the scaling of the dust attenuation in the narrow line region (NLR). 
While the BLR and accretion disk are subject to the nominal $A_V$, the more spatially-extended NLR is attenuated by $\eta_{\rm NLR} A_V$. 
We adopt a fixed $\eta_{\rm NLR} = 0.3$, consistent with the LRD Balmer decrement measurements from \citet{killiDeciphering2023}.

We fit these QSO models using the \textsc{MultiNest} nested sampling algorithm \citep{ferozMultimodal2008, ferozMULTINEST2009, ferozImportance2019}. 
As a test, we apply our SED fitting code to the reported \JWST/NIRCam photometry for the two confirmed $z>7$ broad-line AGN reported in \citet{furtakHigh2024} and \citet{kokorevUNCOVER2023}. 
We recover the spectroscopic redshifts within $1\sigma$, and our estimate of the bolometric luminosity (from the continuum) is consistent with the estimate derived from the H$\beta$ line for the source in \citet{kokorevUNCOVER2023}. 
We note that we overestimate the bolometric luminosity for the source in \citet{furtakJWST2023} by a factor of $\sim 5$, though this is consistent with the continuum-based $L_{\rm bol}$ estimate from \citet{labbeUNCOVER2023}.

\subsubsection{Galaxy models}\label{sec:sedfitting:bagpipes}

To test the other ``edge case,'' that the population is dominated by compact star-forming galaxies, we fit each object to a galaxy SED model. 
Here we assume that all the light originates from the galaxy, including the blue/UV and red/optical components. 
We use \textsc{bagpipes}, a Bayesian SED fitting code \citep{carnallInferring2018a} designed to fit galaxy photometry and spectra. 
We use the BPASS library of stellar SED models \citep{eldridgeBinary2017} and a non-parametric SFH model.
The SFH is parametrized by the $\Delta \log({\rm SFR})$ in adjacent time bins; we adopt the continuity prior \citep[described in][]{lejaHow2019}, i.e.~the prior on $\Delta\log({\rm SFR})$ is a $t$-distribution with $\sigma=0.3$ and $\nu=2$ degrees of freedom. 
We adopt five age bins for the SFH, with the first three at fixed ages from $0$--$10$ Myr, $10$--$50$ Myr, and $50$--$200$ Myr and two more logarithmically-spaced from 200 Myr to $z=20$.

We adopt log-uniform priors on the stellar mass (from $10^{6}$ to $10^{13}\,M_\odot$) and metallicity (from $10^{-3}$ to $1.5\,Z_\odot$). 
Nebular emission is implemented via \textsc{cloudy}, specifically the updated grids presented in \citet{bylerNebular2017}, with $\log U_{\rm ion}$ allowed to vary from $-4$ to $-1$.
We adopt an SMC dust law with $A_V$ allowed to vary from 0 to 6; to be consistent with our QSO modeling, we modify the dust attenuation prescription to include a similar ``scattering fraction'' parameter, which we allow to vary with a log-uniform prior from $10^{-5}$ to $10^{-1}$. 
We note, however, that in the galaxy case, this is better interpreted as representing ``leakage'' through holes in the ISM dust screen, as is found in some lower-redshift DSFGs \citep[e.g.][]{caseyAre2014}.

\subsection{Individual Objects}

We highlight a few individual objects which stand out from our sample. 
Figure~\ref{fig:SEDs1} shows the cutouts and SEDs for six particularly bright LRDs ($m_{444} \sim 22$--$24$). 
In each panel, the best-fit QSO model is shown in black, and the best-fit \textsc{bagpipes} model is shown in blue.
For sources which fall in the PRIMER coverage, we plot the PRIMER photometry as well.  
For the brightest LRDs in our sample, we derive bolometric luminosities $10^{46-47.5}$ erg\,s$^{-1}$ from the QSO modeling, or stellar masses $10^{10-12}$ from the \textsc{bagpipes} modeling. 
This is due to their steep, red optical continuum slopes, anchored by bright F444W/F770W magnitudes, implying significant dust obscuration. 

\begin{figure*}
	\centering
	\includegraphics[width=0.9\linewidth]{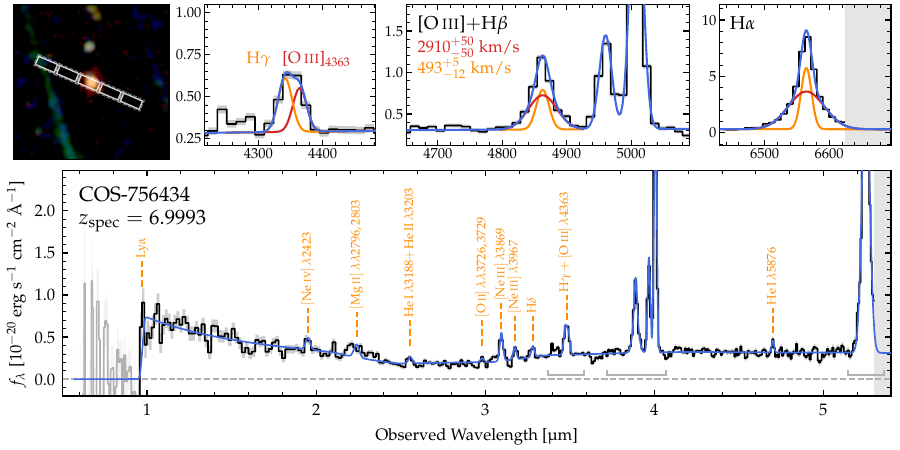}
	\caption{\JWST/NIRSpec PRISM spectrum of COS-756434. The source is clearly detected in continuum and exhibits strong \oiii+H$\beta$ and H$\alpha$ emission, confirming the redshift at $z=6.9993\pm 0.0001$, consistent with the photometric redshift $z_{\rm phot} \sim 7.05^{+0.09}_{-0.05}$. 
	Broad components are visible in both the H$\beta$ and H$\alpha$ lines, with a FWHM of $2750\pm 40$ km\,s$^{-1}$, a clear signature of AGN activity.
	The top panels highlight the broad+narrow line decomposition for H$\beta$ and H$\alpha$, as well as the deblending of \oiii$\lambda 4363$ and H$\gamma$. 
	The source also exhibits several high-ionization emission lines, including [Ne\,\textsc{iv}]$\lambda 2423$ and [Ne\,\textsc{iii}]$\lambda 3869$, consistent with the AGN interpretation.}\label{fig:spec}
\end{figure*}

Several bright LRDs in our sample also fall into the PRIMER MIRI coverage, which includes F1800W. 
We highlight four of these objects in Figure~\ref{fig:SEDs2}. 
The only object in our sample detected in F1800W is COS-105829, which is also the brightest LRD in the F1800W coverage ($m_{770} \sim 22.7$) but is not in the PRIMER NIRCam coverage. 
No other source is detected, despite several (pictured in Figure~\ref{fig:SEDs2}) being similarly bright in F444W and F770W.   
The MIRI limits imply flat SED (in $f_\nu$) between observed-frame $7.7$--$18$\,\textmu m, or rest-frame $\sim 1$--$3$\,\textmu m at the redshifts of these objects.
This may indicate marginal contribution from a hot dust torus, which we return to in Section~\ref{sec:discussion:dust}. 

We provide the SED-derived properties of our sample in Table~\ref{tab:sources}.

%
%Figure~\ref{fig:SEDs3} shows the cutouts and SEDs for four particularly high-redshift $z\gtrsim 7$ candidates. 
%
%
%\red{[Add some more discussion of the individual higher-redshift candidates.]}
%
%Could comment on the relative security of identifications in PRIMER compared to COSMOS-Web, but keep it brief because your paper last year did this at some level

\subsection{Spectroscopic Confirmation of COS-756434}

We identify one source in our sample, COS-756434, with public \JWST/NIRSpec observations. 
COS-756434 was observed with $\sim 6$ hr PRISM spectroscopy as part of a DDT program (\#6585, P.I.~D.~Coulter) targeting high-redshift supernovae.
We reduce the NIRSpec data using the standard \JWST\ pipeline (version 1.14.0); the automatically extracted 1D spectrum is shown in Figure~\ref{fig:spec}. 
The spectrum exhibits strong \oiii\,$\lambda\lambda4959,5007$$+$H$\beta$ and H$\alpha$ emission, confirming the redshift at $z_{\rm spec}=6.9993$, consistent with the photometric redshift of $z_{\rm phot} \sim 7.05^{+0.09}_{-0.05}$. 
We fit the spectrum with a simple model including narrow+broad H$\beta$ and H$\alpha$ emission, narrow emission for other detected lines, and a piecewise power-law continuum. 
We detect the H$\gamma$ and \oiii\,$\lambda4363$ lines, $[{\rm Ne}\,\textsc{iii}]\,\lambda\lambda 3869,3967$ and even the high-ionization line $[{\rm Ne}\,\textsc{iv}]\,\lambda2423$, consistent with the AGN interpretation.
We measure a broad-line FWHM $= 2920^{+40}_{-50}$ km/s, corresponding to a black hole mass $\log M_{\rm BH} \sim 7.8 \pm 0.5$ using the calibration from \citet{greeneEstimating2005}. 
Though we do not discuss the spectrum of this object in detail, we report the properties measured from the spectrum in Table~\ref{tab:spec} and adopt its spectroscopic redshift in the remainder of the paper.

\begin{deluxetable}{@{\extracolsep{10pt}}l@{}c@{}C@{}}
\tabletypesize{\small}
\centering
\tablecaption{Spectroscopic measurements for COS-756434}\label{tab:spec}
\tablehead{\colhead{Property} & \colhead{Units} & \colhead{Value}}
\startdata
$z_{\rm spec}$ & \dots & 6.9993\pm 0.0001 \\
R.A. & hms & 10\mathrm{:}00\mathrm{:}25.6561 \\
Decl. & dms & +02\mathrm{:}21\mathrm{:}36.165 \\
$f_{\lambda,5100}$ & $10^{-21}$\,erg\,s$^{-1}$\,cm$^{-2}$\,\AA$^{-1}$ & 3.20_{-0.03}^{+0.04} \\
$A_{V{\rm ,BLR}}$ & AB mag & 2.9^{+0.2}_{-0.2} \\
$\log M_{\rm BH}$ & $M_{\odot}$ & 7.8 \pm 0.5 \\
$\log L_{{\rm bol,H}\alpha}$ & erg\,s$^{-1}$ & 45.2 \pm 0.4 \\
$\lambda_{\rm Edd}$ & \dots & 0.19^{+0.30}_{-0.11} \\[0.1em]
\hline
FWHM$_{\rm narrow}$ & km\,s$^{-1}$ & 494^{+5}_{-11} \\
FWHM$_{\rm broad}$ & km\,s$^{-1}$ & 2920^{+40}_{-50} \\
$F_{[{\rm O}\,\textsc{iii}]\lambda4363}$ & $10^{-20}$\,erg\,s$^{-1}$\,cm$^{-2}$ & 7.07_{-0.67}^{+0.63} \\
$F_{{\rm H}\beta{\rm ,narrow}}$ & $10^{-19}$\,erg\,s$^{-1}$\,cm$^{-2}$ & 1.18_{-0.13}^{+0.13} \\
$F_{{\rm H}\beta{\rm ,broad}}$ & $10^{-19}$\,erg\,s$^{-1}$\,cm$^{-2}$ & 0.23_{-0.02}^{+0.02}\\
$F_{[{\rm O}\,\textsc{iii}]\lambda4959}$ & $10^{-19}$\,erg\,s$^{-1}$\,cm$^{-2}$ & 2.77_{-0.07}^{+0.07} \\
$F_{[{\rm O}\,\textsc{iii}]\lambda5007}$ & $10^{-19}$\,erg\,s$^{-1}$\,cm$^{-2}$ & 7.98_{-0.08}^{+0.08} \\
$F_{{\rm H}\alpha{\rm ,narrow}}$ & $10^{-18}$\,erg\,s$^{-1}$\,cm$^{-2}$ & 1.11_{-0.03}^{+0.02}  \\
$F_{{\rm H}\alpha{\rm ,broad}}$ & $10^{-18}$\,erg\,s$^{-1}$\,cm$^{-2}$ & 2.44_{-0.03}^{+0.03}  \\
\enddata
\end{deluxetable}

\section{Two Interpretations of LRDs}

We now consider the two alternative interpretations of LRDs as either reddened quasars/dusty AGN or compact star-forming galaxies. 
In the following sections we explore the implications of either scenario, in particular the resulting mass/luminosity functions. 

\subsection{Dusty AGN/reddened quasars}\label{sec:agn}

First, we discuss the interpretation of the LRDs as a population of reddened AGN, as has been posited \citep[e.g.][]{labbeUNCOVER2023, mattheeLittle2024}. 
We characterize the LRDs in terms of their bolometric luminosity, i.e.~the total AGN luminosity after correcting for dust attenuation; we consider the bolometric luminosity rather than the UV luminosity as these sources are quite obscured. 
We compute the bolometric luminosity from the intrinsic model SED (i.e.,~before any dust attenuation), using the monochromatic luminosity at $3000$ \AA\ and a bolometric correction of 5.15 \citep{richardsSpectral2006}. 

\begin{figure}
\centering
\includegraphics[width=\linewidth]{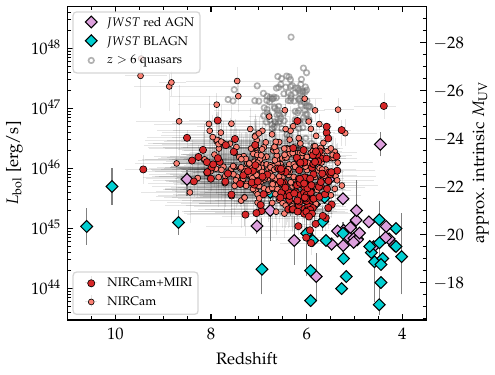}
\caption{Bolometric luminosity vs.~redshift. 
AGN from the literature with \JWST\ spectroscopy are shown in purple \citep[reddened AGN;][]{greeneUNCOVER2024,furtakHigh2024,kokorevUNCOVER2023,mattheeLittle2024} and blue \citep[typical broad-line AGN;][]{larsonCEERS2023,harikaneJWST2023a,maiolinoSmall2024,bogdanEvidence2024,maiolinoJADES2023}. Our sample from COSMOS-Web spans the range of luminosities between the literature \JWST\ AGN and the UV-bright quasars discovered in wide-field ground-based surveys \citep{fanQuasars2023}.
}\label{fig:Lbol_z}
\end{figure}

Figure~\ref{fig:Lbol_z} shows the COSMOS-Web LRDs in the AGN bolometric luminosity vs.~redshift plane. 
Literature AGN with \JWST\ spectroscopy are shown in purple (reddened AGN) and blue (broad-line AGN). 
We also show a compilation of bright quasars from the literature \citep{fanQuasars2023}. 
The COSMOS-Web LRDs span the range between the literature \JWST\ samples and the bright quasars, with a handful reaching similar bolometric luminosities $\sim 10^{47}$ erg\,s$^{-1}$. 
These luminosities are generally much higher than other LRD samples, due to the wide/shallow nature of COSMOS-Web, but we note that similarly luminous LRDs have been observed at $z\sim 3$--$5$ (e.g.~UNCOVER-45924 at $z=4.66$; \citealt{greeneUNCOVER2024}, and RUBIES-BLAGN-1 at $z=3.1$; \citealt{wangRUBIES2024}). 
However, the high luminosities and abundance of these objects is surprising given the relative scarcity of luminous quasars.

To quantify the excess of luminous AGN in our sample, we compute the bolometric luminosity function.
We note that here, and in our estimates of the stellar mass function in  Section~\ref{sec:sfgs}, we compute the volume of the survey as the differential comoving volume scaled to the survey area of 0.54 deg$^2$ and integrated over the redshift bin. 
In order to capture the large uncertainties on the redshifts and luminosities of our objects, we marginalize over the joint posterior distribution of $L_{\rm bol}$ and redshift for each source; that is, individual objects can fractionally span multiple redshift and luminosity bins. 
We additionally marginalize over the bin size, computing the luminosity function in bins centered at $\log L_{\rm bol} \sim 43$, $43.2$, $\dots$, $48$ with bin widths randomly drawn from $0.3$ to $1.0$ dex. 
This avoids biases imparted by the choice of bin size.

We additionally employ a simple completeness correction based on our photometric catalog and source selection procedure. 
In short, we inject 10,000 mock sources, constructed to be compact and extremely red, into the NIRCam data and repeat our cataloging procedure and selection. 
We then estimate the completeness as a function of $m_{444}$, for a color distribution representative of our full sample. 
We translate this to be a function of $L_{\rm bol}$ using the correlation between $m_{444}$ and the median $L_{\rm bol}$ derived from our SED fitting for each source. 
Uncertainty on the completeness is computed from the same correlation using the 16th and 84th percentile $L_{\rm bol}$; this is added in quadrature with the Poisson uncertainty from the luminosity function.

\begin{figure}
\centering
\includegraphics[width=\linewidth]{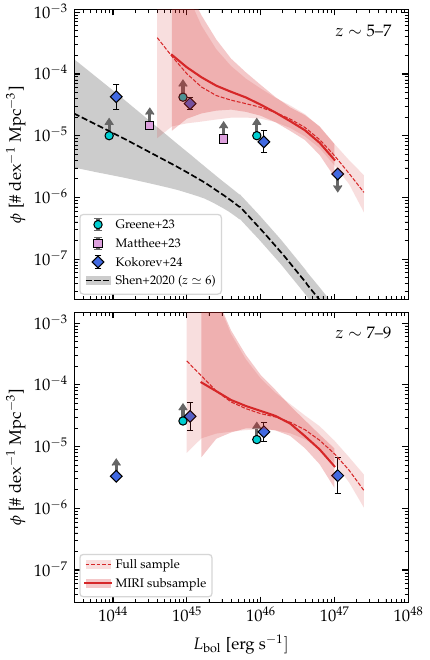}
\caption{AGN bolometric luminosity function at $z\sim 5$--$7$ (top) and $z\sim 7$--$9$ (bottom). Our results are shown in red, for the MIRI subsample (solid) and the full sample (dashed). The bolometric luminosity function limits from \citet{greeneUNCOVER2024} and \citet{mattheeLittle2024}, based on spectroscopically-confirmed LRDs, are shown in purple/blue. Photometric estimates from \citet{kokorevCensus2024}, derived from other \JWST\ imaging surveys, are shown in dark blue. The $z\sim 6$ quasar luminosity function from \citet{shenBolometric2020} is shown in grey. Our results imply a factor of $\sim 100$ overabundance of luminous AGN relative to pre-\JWST\ estimates from UV-bright quasars.}\label{fig:bol_LF}
\end{figure}

\begin{deluxetable}{cCCC}
\centering
\tabletypesize{\small}
\tablecaption{Binned AGN bolometric luminosity function}\label{tab:bolLF}
\tablehead{\multirow{2}{*}{Redshift} & \colhead{$\log L_{\rm bol}$} & \multicolumn{2}{c}{$\log \phi~[\#\,{\rm dex}^{-1}\,{\rm Mpc}^{-3}]$} \\[-0.7em]
& \colhead{$[{\rm erg\,s}^{-1}]$} & \colhead{Full sample} & \colhead{MIRI subsample} }
\startdata
5--7 & 45 & -3.99^{+1.31}_{-0.81} & -3.89^{+1.33}_{-0.84} \\
5--7 & 45.5 & -4.39^{+0.65}_{-0.34} & -4.24^{+0.64}_{-0.35} \\
5--7 & 46 & -4.54^{+0.31}_{-0.18} & -4.50^{+0.32}_{-0.18} \\
5--7 & 46.5 & -4.80^{+0.22}_{-0.14} & -4.84^{+0.23}_{-0.16} \\
5--7 & 47 & -5.32^{+0.19}_{-0.16} & -5.39^{+0.25}_{-0.28} \\
\hline
7--9 & 45.5 & -4.19^{+1.30}_{-0.59} & -4.17^{+1.33}_{-0.65} \\
7--9 & 46 & -4.42^{+0.58}_{-0.28} & -4.42^{+0.58}_{-0.28} \\
7--9 & 46.5 & -4.67^{+0.26}_{-0.17} & -4.73^{+0.27}_{-0.20} \\
7--9 & 47 & -5.32^{+0.29}_{-0.35} & -5.32^{+0.29}_{-0.35} \\
\enddata
\end{deluxetable}

\begin{figure}
\centering
\includegraphics[width=\linewidth]{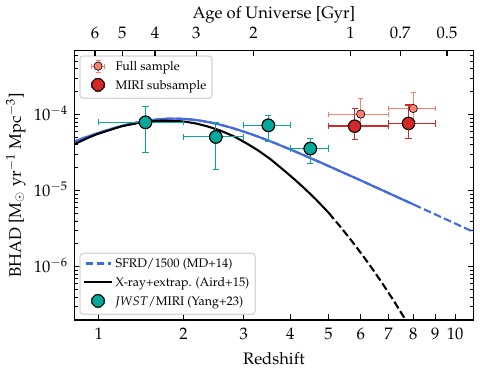}
\caption{The black hole accretion rate density as a function of redshift. Our results are shown in red, for our full sample (light red, smaller points) and our MIRI subsample (dark red, larger points). 
Results from X-ray AGN surveys \citep{airdEvolution2015} are shown in black, including extrapolation at $z>5$. 
Results from \JWST/MIRI are shown in green \citep{yangCEERS2023}.
If the red dots are universally AGN, our results imply a non-evolving black hole accretion density of $\sim 0.5$--$1\times 10^{-4}$ M$_\odot$ yr$^{-1}$ Mpc$^{-3}$ from $z\sim 1$ to $z\sim 9$.}	\label{fig:BHAD}
\end{figure}

Figure~\ref{fig:bol_LF} shows the AGN bolometric luminosity function inferred from our photometric candidate sample at $z\sim 5$--$7$ (top) and $z\sim 7$--$9$ (bottom). 
Our sample is shown in red, with the solid line corresponding to the MIRI subsample. 
We also show lower limits on the bolometric LF from LRDs selected in spectroscopic surveys (UNCOVER, \citealt{greeneUNCOVER2024}, and FRESCO, \citealt{mattheeLittle2024}) and estimates from other photometric surveys \citep{kokorevCensus2024}. 
Our results are consistent with spectroscopic samples, and suggest an overall  excess of red AGN relative to the quasar bolometric luminosity function measured from rest-UV selected quasars (shown in grey, \citealt{shenBolometric2020}).
While the faint-end slope of the quasar bolometric luminosity function remains uncertain, our results suggest that the knee of the LF could be dominated by these dust-reddened AGN, even up to $L_{\rm bol} \sim 10^{47}$ erg\,s$^{-1}$. 
In particular, we find a factor of $\sim 100$ over-abundance of the LRDs relative to UV-bright quasars, at fixed $L_{\rm bol}$. 
This could also be explained if we were overestimating the bolometric luminosities of the LRDs, though we note that this would also require a factor of $\sim 100$ in luminosity to be consistent with the QLF. 
Our estimated bolometric luminosity function is tabulated in 0.5 dex bins of luminosity in Table~\ref{tab:bolLF}

How does the abundance of AGN at $z\gtrsim 5$ compare to the population at lower redshift? 
Figure~\ref{fig:BHAD} shows the evolution of the black hole accretion rate density (BHAD) with redshift. 
Results from X-ray AGN surveys \citep{airdEvolution2015} are shown in black, extrapolated above $z\sim 5$. 
We also show the SFRD \citep[scaled down by $1500$;][]{madauCosmic2014} as black-hole galaxy co-evolution scenarios would suggest that the BHAD should follow a similar evolution to the SFRD. 
Results from \JWST/MIRI observations of $z\sim 1$--$5$ obscured AGN in CEERS are shown in blue \citep{yangCEERS2023}. 
We compute the BHAD by integrating the bolometric luminosity function and assuming a radiative efficiency factor $\epsilon = 0.1$.
As Figure~\ref{fig:BHAD} shows, if the LRD are universally AGN, our results imply a non-evolving black hole accretion density of $\sim  10^{-4}$ M$_\odot$ yr$^{-1}$ Mpc$^{-3}$ from $z\sim 1$ to $z\sim 9$.
This is an order of magnitude above the scaled SFRD, and several orders of magnitude above the X-ray extrapolations.

\subsection{Compact SFGs}\label{sec:sfgs}

Next we consider the interpretation of the ``little red dots'' as compact dust-obscured star-forming galaxies, as has been suggested in the literature \citep[e.g.][]{labbePopulation2023, akinsTwo2023, williamsGalaxies2023, perez-gonzalezWhat2024}. 
Under this interpretation, the continuum emission would originate primarily from starlight, with the red color either coming from an old stellar population with a strong Balmer break or high EW emission lines combined with strong dust attenuation. 
The lack of mid-IR detections for these sources may further support their star-forming nature \citep{williamsGalaxies2023,perez-gonzalezWhat2024}, and the relative abundance is more consistent with galaxies than quasars.

\begin{figure}
\centering
\includegraphics[width=\linewidth]{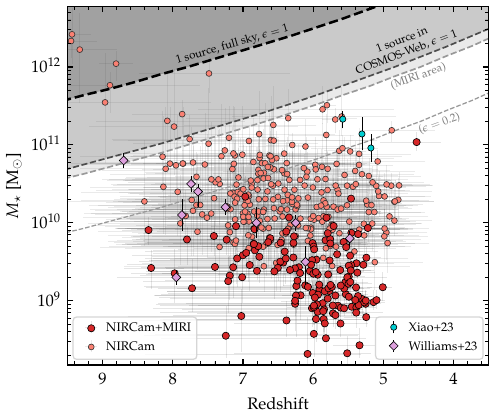}
\caption{Stellar mass vs.~redshift from galaxy SED fitting with \texttt{bagpipes}. 
The dashed lines indicate the maximum stellar mass we would expect to find in a given volume based on the halo mass function (assuming a global star-formation efficiency $\epsilon=1$, or $\epsilon=0.2$ for a more realistic assumption). 
We plot the stellar masses inferred for ``little red dots'' in JADES from \citet{williamsGalaxies2023}, as well as the three ultra-massive objects identified in \citet{xiaoMassive2023}.}\label{fig:mstar_z}
\end{figure}

%As described in Section~\ref{sec:sedfitting:bagpipes}, we fit each object to a galaxy SED model using \texttt{bagpipes} \citep{carnallInferring2018}. 
Figure~\ref{fig:mstar_z} shows the derived stellar mass vs.~redshift for our LRD sample. 
We indicate the region of parameter space that would be disfavored or disallowed in a $\Lambda$CDM framework in dark grey, corresponding to the stellar mass of the most massive halo in a given area assuming the cosmic baryon fraction of $f_b = 0.158$ and stellar baryon fraction $\epsilon = 1$.
Here we compute the halo mass function using the \texttt{hmf} python package \citep{murrayHMFcalc2013} assuming an SMT fitting function \citep{shethEllipsoidal2001}.
We plot the relevant threshold for the full sky, the COSMOS-Web area ($0.5$ deg$^2$) and the MIRI area ($0.17$ deg$^2$) in successively lighter shades of grey. 
We note that a more reasonable upper limit on the observed stellar mass would use $\epsilon \approx 0.2$, the maximum inferred from the peak of the stellar mass to halo mass relation across a range of redshifts \citep{shankarNew2006,mandelbaumGalaxy2006,conroyConnecting2009,behrooziComprehensive2010,behrooziUNIVERSEMACHINE2019,shuntovCOSMOS20202022}.

A few of our LRDs exceed the maximum thresholds expected/allowed in $\Lambda$CDM, particularly at $z\gtrsim 7$. 
This may be seen as reason to favor the AGN interpretation: if they are dominated by stellar light, they are simply too massive.  
However, we note that the subsample with MIRI coverage does not violate these thresholds; the stellar masses derived for the MIRI subsample are generally lower than those for the NIRCam-only sample. 
To test this further, we perform a series of \texttt{bagpipes} runs on the MIRI sample, both including and omitting the F770W data point. 
We find systematic offsets of $\sim 1$--$2$ dex, towards larger masses, when not including the MIRI data. 
Similar results have been found in other surveys with MIRI imaging (e.g.~CEERS, \citealt{papovichCEERS2023}; SMILES, \citealt{williamsGalaxies2023}). 
In our case, this effect is largely due to the impact of strong emission lines boosting the photometry in the F444W bands. 
With the additional data point on the red end, the modeling tends to favor younger stellar populations with stronger emission lines and lower masses---i.e., a compact starburst. 
We note, however, that the total contribution of the LRDs to the cosmic star-formation rate density is still $\lesssim 5\%$, based on the sample with MIRI data.  
We therefore conclude that the LRDs do not individually violate $\Lambda$CDM limits, but that MIRI imaging is critical to derive accurate masses, especially in the case of limited broadband coverage.

\begin{figure}
\centering
\includegraphics[width=\linewidth]{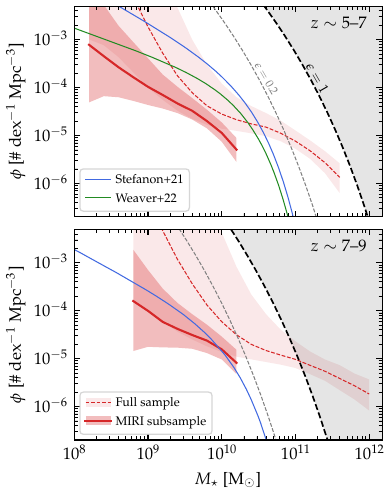}
\caption{Stellar mass functions at $z\sim 5$--$7$ (top) and $z\sim 7$--$9$ (bottom). As in Figure~\ref{fig:bol_LF}, our sample is shown in red, for the MIRI subsample (solid) and the full sample (dashed). The blue and green lines show the pre-\JWST\ estimates of the SMF from \citet{stefanonGalaxy2021a} and \citet{weaverCOSMOS20202023}. The grey shaded region shows the parameter space prohibited by $\Lambda$CDM, scaling the the halo mass function by the cosmic baryon fraction $f_b$ and a maximal ``efficiency'' $\epsilon=1$. The grey dashed line shows a more realistic efficiency $\epsilon=0.2$. While the SMF for the full sample violates the limits imposed by $\Lambda$CDM, particularly at $z>7$, the impact of MIRI on the stellar masses brings the SMF into agreement with literature estimates.}
\label{fig:SMF}
\end{figure}

However, even without individual objects violating $\Lambda$CDM, the ubiquity of these sources suggests a high number density of massive galaxies. 
Figure~\ref{fig:SMF} shows the stellar mass function (SMF) derived from our LRD sample. 
We show the pre-\JWST\ estimates of the SMF from CANDELS \citep{stefanonGalaxy2021a} and COSMOS \citep{weaverCOSMOS20202023} in blue and green, respectively. 
As discussed previously, our full sample shows a strong excess relative to pre-\JWST\ estimates, and with several candidates in the region prohibited by $\Lambda$CDM. 
However, the subsample with MIRI coverage does not show as strong of an excess, and in fact shows a bright-end slope more consistent with the steep slope derived in previous studies. 
This suggests that these objects could be dominated by star-formation. 
Importantly, though, this implies that the massive galaxy population at $z\gtrsim 5$ is dominated by these compact starbursts, which has important implications for our understanding of early stellar mass growth.

\begin{figure}
\centering
\includegraphics[width=\linewidth]{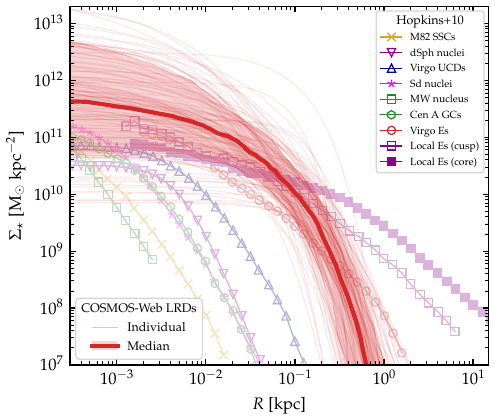}
\caption{Median stellar mass surface density profile for the COSMOS-Web LRDs, compared to the compilation of dense stellar systems in \citet{hopkinsMaximum2010}. Individual profiles are shown in the thin, light red lines. The different colors/symbols correspond to the different classes of objects compiled by \citet{hopkinsMaximum2010}, ranging from local globular clusters/nuclear disks to massive ellipticals at $z\gtrsim 2$; all exhibit similar central densities, roughly consistent with the median LRD in our sample.}\label{fig:sigmastar_profiles}
\end{figure}

A stronger constraint can be placed on the stellar content of these objects based on their point-like morphology. 
There exists a maximum stellar mass surface density observed in dense stellar systems \citep{hopkinsMaximum2010}, which remains consistent across $z\sim 0$--$3$ and over $\sim 9$ dex in stellar mass. 
In other words, local and $z\lesssim 3$ stellar systems show similar maximum stellar densities $\sim 10^{11}$ M$_\odot$ kpc$^{-2}$ in their cores. 
Figure~\ref{fig:sigmastar_profiles} shows the stellar mass profiles derived for our sample, in comparison to the average profiles for various local and high-redshift populations from \citet{hopkinsMaximum2010}. 
Here we include only the MIRI subsample, for which we can derive reliable stellar masses. 
We find a median central stellar density of $\log \Sigma_\star \sim 11.6$, higher than---but broadly consistent with---that observed in the lower-redshift universe. 
Approximately half of the sample, however, has stellar mass profiles significantly in excess of this limit. 
On the one hand, these central densities may be reduced over time due to $N$-body relaxation, the timescale for which can be $\lesssim 1$ Gyr for particularly compact stellar systems \citep{hopkinsMaximum2010}. 
On the other hand, this may imply that the stellar masses are simply overestimated, requiring significant AGN contribution to the continuum to reconcile the derived central densities.

\section{Multiwavelength Stacking}\label{sec:stacking}

\begin{figure*}
\centering
\includegraphics[width=\linewidth]{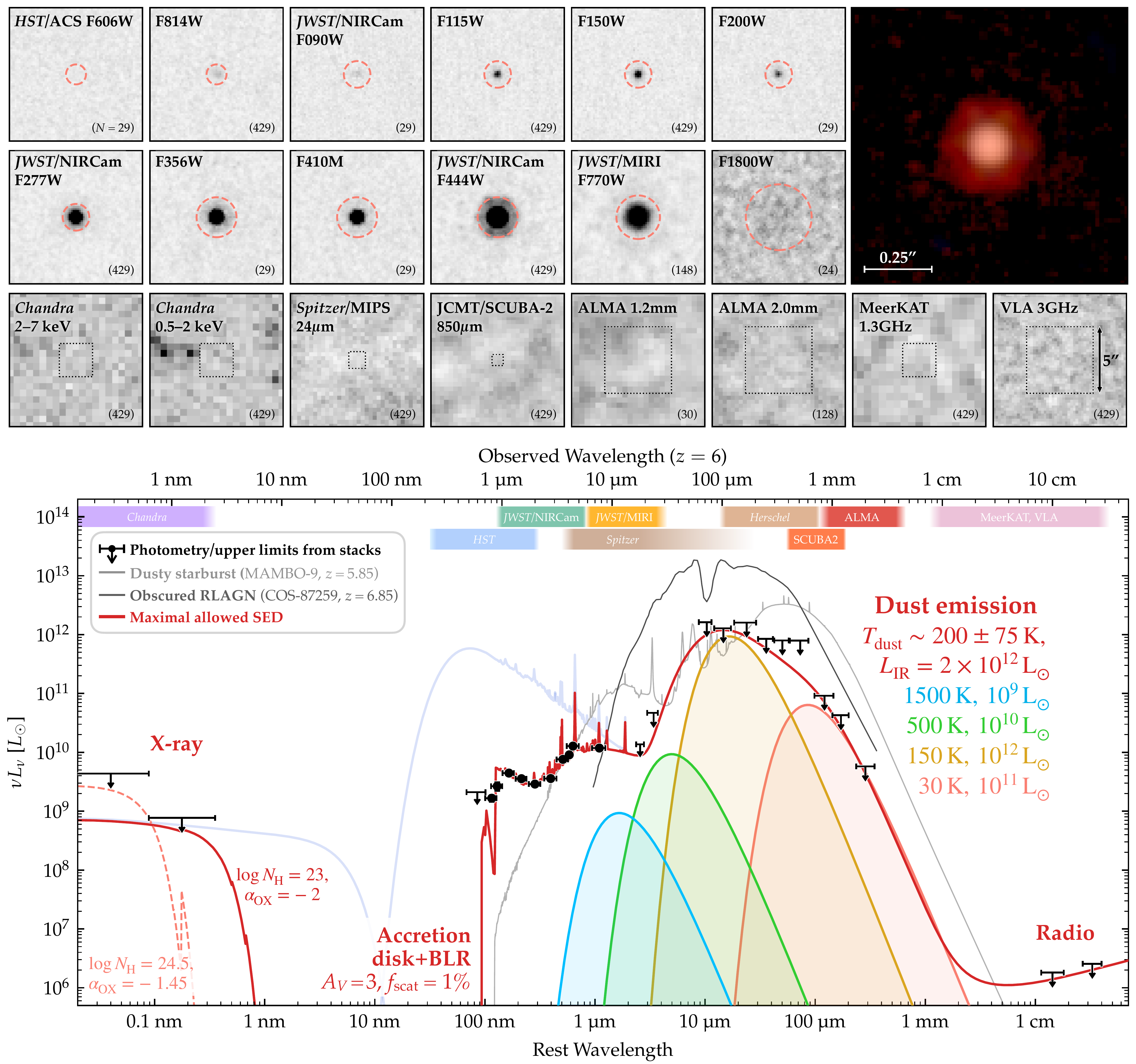}
\caption{Median-stacked panchromatic SED for the COSMOS-Web LRDs. \textit{Top:} Stacked images in all \HST/\JWST\ bands available, as well as \textit{Chandra} X-ray, \textit{Spitzer}/MIPS 24\,\textmu m, JCMT/SCUBA-2 850\,\textmu m, ALMA 1.2 and 2.0 mm imaging, and MeerKAT/VLA 1.3 and 3GHz imaging. Each stack is annotated with the number of objects contributing to the stack. For \HST/\JWST, we show in light red the apertures used to extract photometry directly from the stacks. For the remaining bands, we show a 5'' square to guide the eye given the varying resolution of the images. 
\textit{Bottom:} full panchromatic SED for the median COSMOS-Web LRD. Photometry measured from stacks (or $5\sigma$ upper limits from the non-detections) are shown in black. 
We plot the maximal allowed SED model in red. 
The faint blue lines in the X-ray and UV/optical show the intrinsic (i.e., unattenuated) SEDs for the corona and the accretion disk.
We plot two alternative X-ray SEDs consistent with the data, a Compton-thin model with an anomalously low $\alpha_{\rm OX}$ (solid) and a Compton-thick model with a normal $\alpha_{\rm OX}$ given the optical luminosity (dashed). 
In the far-IR, we show several single-temperature blackbodies at different normalizations, ranging from to cold ISM dust the hottest possible dust, at the sublimation temperature. 
The maximal SED model includes a total dust luminosity of $2\times 10^{12}\,L_{\odot}$ with a Gaussian temperature distribution $\sim 200\pm 75$\,K. 
Finally, in the radio, we adopt a slope $\alpha=-1.5$ consistent with radio-loud AGN \citep{endsleyRadio2022, lambridesUncovering2023}.
For reference, we also plot the best-fit model SEDs for a $z=5.85$ DSFG (MAMBO-9; Akins et al.~\textit{in prep}) and a $z=6.85$ obscured AGN \citep[COS-87259;][]{endsleyRadio2022,endsleyALMA2022}.  }\label{fig:stack}
%Based on the optical/X-ray SED slope $\alpha_{\rm OX}$, the stacked X-ray imaging \textit{should} yield a detection, even with the moderate column density ($10^{23}$ cm$^2$) expected based on the dust-reddening ($A_V=3.5$); see text for details. 
%At the other end of the spectrum, the \textit{Spitzer}/MIPS 24\,\textmu m stack provides a limit on the hottest dust in the AGN torus, at the sublimation temperature ($T\sim 1500$ K).  
%Similarly, the SCUBA-2 and ALMA stacks provide limits on the cold dust component ($T\sim 30-60$ K) heated by star-formation. 
%The typical little red dot has $\sim \sfrac{1}{50}^{\rm th}$ the hot dust luminosity as COS-87259, and $\sim \sfrac{1}{10}^{\rm th}$ the cold dust luminosity as MAMBO-9. 
%No strong constraints can be placed on the contribution from warm ($T\sim 200$ K) dust, which may be associated with a compact nuclear starburst.}	
\end{figure*}
 
The COSMOS field is rich in multiwavelength coverage from the X-ray, MIR, FIR/submm, and radio.
As described in Section~\ref{sec:data}, we inspect cutouts around each source in \textit{Chandra} X-ray imaging, \textit{Spitzer}/MIPS 24\,\textmu m, SCUBA-2 and ALMA mm/sub-mm imaging, and VLA/MeerKAT radio data. 
No source in our sample is individually detected in any of these data; this is somewhat surprising, given the large sample size and inclusion of several remarkably luminous objects in the sample. 
In particular, under the AGN interpretation, objects with $L_{\rm bol} \gtrsim 10^{46}$ erg/s at $z\gtrsim 4$ are regularly detected in the X-ray \citep{airdEvolution2015}, mid-infrared (from the dusty torus), and radio \citep[e.g.][]{endsleyALMA2022,lambridesUncovering2023}.

To push the multiwavelength constraints deeper, we stack the data for the entire sample.
Even without individual detections, if these objects exhibit ubiquitous, faint flux, it would show up in stacking. 
We perform a simple median stack of the cutouts in each map. 
We note that for \textit{Chandra} we compute the \textit{sum}, rather than the median, as the median is always 0 given the low X-ray counts. 
We additionally perform stacking on \textit{Spitzer}/MIPS 70\,\textmu m and \textit{Herschel} imaging from PACS \citep[100 \& 160\,\textmu m;][]{lutzPACS2011} and SPIRE \citep[250, 350, and 500\,\textmu m][]{oliverHerschel2012}; while SPIRE imaging is severely confusion-limited, stacking on maps with a mode equal to 0 removes flux boosting from adjacent sources \citep[see also][]{coppinSCUBA22015}.
Due to the widely varying PSF size of the various datasets, we stack relatively large cutouts (ranging from $5''$ for \HST/\JWST\ to $10'$ for \textit{Herschel}/SCUBA data).

The stacked cutouts are shown in the top panels of Figure~\ref{fig:stack}. 
No flux is detected in our stacks in any band aside from \HST/\JWST.
We measure fluxes directly from the stacked cutouts for \HST+\JWST, yielding the median (observed-frame) SED for the sample of LRDs; note that we find marginal flux in F1800W ($2.9\sigma$), but formally take this as an upper limit.

To quantify the flux limits implied by the non-detections in the various multi-wavelength datasets, we plot in the lower panel of Figure~\ref{fig:stack} the median panchromatic SED for the COSMOS-Web LRDs. 
We derive the upper limits for each band in the following manner. 
For the ALMA data, which comes from the CHAMPS (Faisst et al.~\textit{in prep}) and MORA surveys \citep[][Long et al.~\textit{in prep}]{caseyMapping2021}, and SCUBA-2, we add in quadrature the RMS at the position of each source, and divide by $N$. 
For all other maps, which have relatively uniform depth, we adopt the reported depths, scaled down by $\sqrt{N}$. 
For \textit{Chandra}, we use the limiting depths reported in \citet{civanoChandra2016} for the $0.5$--$2$ and $2$--$7$ keV bands. 
For \textit{Spitzer}/MIPS and \textit{Herschel}/PACS and SPIRE we use the depths reported in \citet{sandersSCOSMOS2007}, \citet{lutzPACS2011}, and \citet[][confusion limits]{nguyenHerMES2010}, respectively.

Do the stacked limits provide useful constraints on the physical properties of the LRDs? 
To answer this, we show in the dark red line in Figure~\ref{fig:stack} the \textit{maximal} SED allowed by the limits. 
The UV/optical/NIR SED is constructed from our QSO model (see \S\ref{sec:sedfitting:qso}) to match the photometry, with $A_V=3$ and $f_{\rm scat} = 1\%$ at $z=6$.
The intrinsic (unattenuated) SED is shown in light blue, with a UV magnitude of $M_{\rm UV} = -21.8$ and a 3000\,\AA\ continuum luminosity of $L_{3000} = 5.5\times10^{44}$ erg/s. 
Note that these properties are very similar to the average SED-derived properties for the sample as a whole.

For the X-ray SED, we adopt a photon index of $\Gamma=1.8$ and an upper energy cut at 300 keV \citep[following X-CIGALE;][]{yangXCIGALE2020}. 
The normalization is set by the (unattenuated) optical SED, using the conversion parameter $\alpha_{\rm OX} \equiv -0.3838 \log (L_{\nu,2500}/L_{\rm 2\,keV})$. 
We then attenuate the SED assuming a column density $N_{\rm H}$, using the photoelectric absorption cross sections provided in \citet{morrisonInterstellar1983}. 
We show in the darker, solid line the X-ray SED assuming Compton-thin absorption $\log N_{\rm H}/{\rm cm}^{-2} = 23$. 
This column density is consistent with the moderate dust attenuation ($A_V\approx 3$) based on empirical relations between dust attenuation and gas absorption in red quasars \citep{glikmanAccretion2024}. 
With this column density, the X-ray limits require $\alpha_{\rm OX} < -2$; by contrast, we expect $\alpha_{\rm OX}\approx -1.45$ ($\pm 0.1$) from the $\alpha_{\rm OX}$--$L_{2500}$ relation \citep{justXRay2007}. 
An alternative model adopts $\alpha_{\rm OX} = -1.45$ but a much higher column density $\log N_{\rm H}/{\rm cm}^{-2} > 24.5$. 
In this Compton-thick model, the X-ray SED falls just below the limits for the shallower hard-band, and is entirely attenuated in the soft band.
Several other authors have reported X-ray measurements for LRDs that imply lower masses/luminosities than the broad lines in the optical \citep[e.g.][]{yueStacking2024, anannaXray2024,maiolinoJWST2024}, which may indicate a fundamental difference in the nuclear structure in these objects.

Turning to the far-IR, the stacked limits provide a constraint on the dust SED of the LRDs. 
We plot four single-temperature modified blackbody models, from cold ISM dust ($T=30$ K) to the hottest dust in the AGN torus, at the sublimation temperature ($T=1500$ K). 
The ALMA/SCUBA constraints limit the cold dust component to $\lesssim 10^{11}\,$L$_\odot$, and the MIRI/MIPS constraints limit the hottest dust components to $\lesssim 10^{10}$\,L$_\odot$. 
The shallow \textit{Herschel} limits, however, provide little constraints on warm dust $\sim 100$--$200$ K. 
We include in the dark red model SED a dust component with a Gaussian distribution of temperatures, with a mean of 200 K and a standard deviation of 75 K. 
We do not impose any energy balance with the rest-frame optical, instead adopting a luminosity of $2\times 10^{12}$\,L$_\odot$, which falls just below the stacked $5\sigma$ limits from Herschel. 
With the available limits, in the maximal SED model, the vast majority of the bolometric luminosity could be emitted in the far-IR and still result in non-detections.  
We note that the bolometric luminosity inferred from $L_{3000}$, $L_{\rm bol} \approx 7.4\times 10^{11}$\,L$_\odot$ \citep[using a bolometric correction of 5.15;][]{richardsSpectral2006} is still less than the maximal luminosity allowed in the FIR, though only by a factor of $\sim 3$---the current FIR limits are consistent with a standard bolometric correction for LRDs.

Finally, in the radio, we adopt a power-law slope of $\alpha = -1.5$, consistent with radio-loud AGN at high-redshift \citep{endsleyRadio2022,lambridesUncovering2023}. 
The normalization is set by the luminosity density at 1.4 GHz, $L_{\nu,1.4\,{\rm GHz}} = 5.2\times 10^{23}$ W\,Hz$^{-1}$. 
This is quite a bit lower than we would expect for radio-loud AGN; in terms of the ratio of the ratio of the 1.4 GHz luminosity to the optical $B$-band luminosity, $R^*_{1.4} = L_{\nu,1.4\,{\rm GHz}}/L_{\nu,4400} = 30$ defines the boundary between radio-loud and radio-quiet AGN \citep{cirasuoloThere2003, cirasuoloRadio2006, wangMillimeter2007}. 
Our radio limits constrain $R^*_{1.4}<10$, at the $5\sigma$ level, suggesting that the average LRD is inconsistent with a radio-loud AGN. \\[-1.1cm]

\begin{deluxetable*}{@{\extracolsep{10pt}}l@{}c@{}C@{}c@{}C@{}}
\tabletypesize{\small}
\centering
\tablecaption{Observed-frame stacking results}\label{tab:stacking}
\tablehead{Band & \colhead{Wavelength} & \colhead{$N$} & \colhead{Units} & \colhead{$F_\nu$}}
\startdata
\textit{Chandra} 2--10 keV$^\dagger$ & 0.28 nm & 434 & erg\,s$^{-1}$\,cm$^{-2}$ & $<7.2\times 10^{-17}$ \\
\textit{Chandra} 0.5--2 keV$^\dagger$ & 1.24 nm & 434 & erg\,s$^{-1}$\,cm$^{-2}$ & $<1.1\times 10^{-17}$ \\
    \HST/ACS F606W & 0.60\,\textmu m & 29 & nJy & $<3.85$ \\
    \HST/ACS F814W & 0.81\,\textmu m & 434 & nJy & 4.11 \pm 0.52 \\
	\JWST/NIRCam F090W & 0.90\,\textmu m & 29 & nJy & 7.37 \pm 1.22 \\
	\JWST/NIRCam F115W & 1.16\,\textmu m & 434 & nJy & 15.76 \pm 0.62 \\
	\JWST/NIRCam F150W &  1.50\,\textmu m & 434 & nJy & 16.46 \pm 0.51 \\
	\JWST/NIRCam F200W & 2.00\,\textmu m & 29 & nJy & 17.66 \pm 0.77 \\
	\JWST/NIRCam F277W & 2.77\,\textmu m & 434 & nJy & 30.51 \pm 0.23 \\
	\JWST/NIRCam F356W &  3.58\,\textmu m & 29 & nJy & 83.30 \pm 0.56 \\
	\JWST/NIRCam F410M &  4.08\,\textmu m & 29 & nJy & 113.56 \pm 1.03 \\
	\JWST/NIRCam F444W & 4.42\,\textmu m & 434 & nJy & 173.00 \pm 0.33 \\
	\JWST/MIRI F770W &  7.66\,\textmu m & 148 & nJy & 277.99 \pm 2.94 \\
	\JWST/MIRI F1800W & 18.00\,\textmu m & 24 & nJy & (435.2 \pm 150.9) \\
     \textit{Spitzer}/MIPS & 24\,\textmu m & 434 & \textmu Jy & $<3.43 $ \\
     \textit{Spitzer}/MIPS & 70\,\textmu m & 434 & mJy & $<0.36 $ \\
    \textit{Herschel}/PACS & 100\,\textmu m & 434 & mJy & $<0.40$ \\
    \textit{Herschel}/PACS & 160\,\textmu m & 434 & mJy & $<0.82$ \\
   \textit{Herschel}/SPIRE & 250\,\textmu m & 434 & mJy & $<0.65$ \\
   \textit{Herschel}/SPIRE & 350\,\textmu m & 434 & mJy & $<0.86$ \\
   \textit{Herschel}/SPIRE & 500\,\textmu m & 434 & mJy & $<1.24$ \\
          SCUBA-2 & 850\,\textmu m & 434 & \textmu Jy & $<239$ \\
             ALMA & 1.2\,mm & 30 & \textmu Jy & $<157$ \\
             ALMA & 2.0\,mm & 128 & \textmu Jy & $<35.4 $ \\
         VLA 3\,GHz & 10\,cm & 434 & uJy & $<0.56 $ \\
  MeerKAT 1.28\,GHz & 23\,cm & 434 & uJy & $<1.80$ 
\enddata
\tablenotetext{}{$^\dagger$For \textit{Chandra}, we quote the absolute flux limit (rather than flux density) to be consistent with X-ray surveys. Uncertainties are quoted as $1\sigma$; for non-detections, we quote the $5\sigma$ upper limits.}
\end{deluxetable*}

\section{Discussion}\label{sec:discussion}

Our results do not necessarily favor one interpretation over the other for the origin of the LRDs, and present challenges to our understanding of early galaxy/SMBH growth in either case. 
Under the AGN interpretation, our estimate of the bolometric luminosity function is consistent with the lower limits provided by recent spectroscopic observations, with a $\sim 2$ dex excess relative to blue quasars across $\log L_{\rm bol}\sim 45$--$47$.
Under the galaxy interpretation, our stellar mass function estimate is consistent with $\Lambda$CDM limits, but implies that the LRDs would dominate the high-mass end of the SMF, and have central stellar mass densities significantly greater than the maximum observed at lower-redshift. 
In reality, neither interpretation is likely 100\% true; both galaxy and AGN components will contribute to the SED. 
Moreover, the mechanisms that trigger compact nuclear starbursts (i.e.~those in local LIRGs or URLIRGs) are thought to be the same mechanisms that trigger quasars; specifically, gas-rich major mergers or disk instabilities \citep[e.g.][]{hopkinsCosmological2008}.
Distinguishing between these two possibilities will inevitably require more data, both to examine sub-populations and decompose individual SEDs.

\subsection{What dominates the continuum?}\label{sec:continuum}

%Another scenario to explain the mid-IR observations without invoking a hot dust deficiency is simply that we are overestimating the AGN power based on the rest-optical emission. 
The major remaining uncertainty surrounding the LRDs is whether the red optical continuum emission is dominated by stellar light or the AGN accretion disk. 
This directly impacts the measured stellar masses and AGN bolometric luminosities, and, to a lesser degree, the black hole masses. 
On the one hand, the broad line EWs measured for LRDs with \JWST\ spectroscopy have been found to be largely consistent with local calibrations, i.e.~the continuum is consistent with being dominated by the accretion disk \citep{greeneUNCOVER2024}. 
However, it is possible that the broad lines are unusually strong relative to the accretion disk continuum in these objects. 
This could be due to different AGN physics (such as a higher BLR covering factor, e.g.~\citealt{maiolinoJWST2024}), differential attenuation between a spatially-extended BLR and the accretion disk \citep[as has been suggested for red quasars and DOGs, e.g.][]{hamannExtremely2017, noboriguchiExtreme2022}, or kinematically broadened lines in a multiphase ISM, even without an AGN (e.g. N.~Roy et al.~\textit{in prep}). 
While the likelihood of these scenarios is unclear, and will need to be examined in more depth in future work, each would imply that \textit{the rest-optical continuum emission is not necessarily dominated by the accretion disk}, resulting in lower bolometric luminosity estimates. 
Lower AGN luminosities would be more consistent with the mid-IR limits from MIRI imaging surveys \citep{williamsGalaxies2023,perez-gonzalezWhat2024} and X-ray limits from \textit{Chandra} \citep{yueStacking2024, anannaXray2024}. 
This may also bring the luminosity function into closer agreement with the quasar luminosity function, though, as we note in Section~\ref{sec:agn}, it would require a factor of $\sim 100$ reduction in order to match the bolometric luminosity function from UV-selected quasars. 
The bolometric LF could also come down if the bolometric correction were found to be lower than the nominal value, e.g.~due to the lack of hot dust in this population. 
We note, though, that the the current FIR data are not sufficient to constrain this; our stack gives an upper limit on the 3000\,\AA\ bolometric correction of $\lesssim 15$, entirely consistent with the nominal value of 5.15 \citep{richardsSpectral2006}

While reducing the AGN contribution to the continuum may alleviate the tension with AGN models and the quasar LF, if the rest-optical continuum is instead dominated by stellar light, it implies a significant contribution of these objects to the stellar mass function. 
As we show in Section~\ref{sec:sfgs}, the LRDs could dominate the stellar mass function at the high-mass end, particularly at $z>7$. 
This is true even with the MIRI data generally preferring solutions in which a recent, young starburst dominates the mass, which effectively represent a lower-limits on the mass. 
Recent \JWST/NIRSpec results have identified LRDs with clear Balmer breaks, indicating the presence of old stellar populations alongside broad emission lines \citep{wangRUBIES2024a}; if ubiquitous in LRDs, this would raise their contribution to the stellar mass function. 
However, even with high SNR spectroscopy from $1$--$5$\,\textmu m, the relative contribution from a Balmer break as opposed to a reddened power-law continuum is degenerate, with derived stellar masses ranging from $10^9$ to $10^{11}~M_\odot$ depending on the breakdown of the two components \citep{wangRUBIES2024a}. 
MIRI imaging may help to break some of these degeneracies, as we find in our sample, and will be crucial to carefully decompose the continuum contribution from galaxies/AGN in the future.

\subsection{What is the origin of the dust obscuration?}\label{sec:discussion:dust}

Whether the LRDs represent star-forming galaxies or AGN, their dust-reddening provides a view into the physics of AGN/ISM dust at high-$z$. 
Despite SED-derived attenuation $A_V\gtrsim 2$, the LRDs entirely lack submillimeter or millimeter detections with SCUBA, ALMA, or otherwise, even in our deep stacks of hundreds of objects.  
Their sub-mm faintness has been interpreted as evidence in favor of an AGN interpretation, as the ALMA limits are inconsistent with energy-balance models in which the attenuated starlight is reprocessed in the sub-mm \citep{labbeUNCOVER2023}. 
However, in a given bandpass, sub-mm observations are extremely sensitive to the temperature ($S_\nu \propto T^{-3.5}$ for fixed $L_{\rm IR}$), such that the difference between $T=25$ K and $T=75$ K represents $\sim 100$ times fainter emission \citep[see][]{caseyDusty2014}. 
The typical assumption of a cold temperature ($T\sim 25$--$35$ K) may not apply in these systems; in fact, hotter dust is in fact expected at high redshift, as the increased CMB temperature, $T_{\rm CMB} = 2.73\,{\rm K}\times(1+z)$, sets an absolute floor on the dust temperature. 
Moreover, a simple application of the Stefan-Boltzmann law ($L_{\rm IR} = 4\pi R_e^2\sigma T^4$) would predict higher dust temperatures for a more compact IR-emitting region \citep[see e.g.][]{burnhamPhysical2021}. 
As we show in Figure~\ref{fig:stack}, even a substantial luminosity of warm dust ($T\sim 150$ K) would be completely missed at the current limits of our samples, but would be expected if the obscuration originates on small scales.
This warm dust could be heated by the AGN, either directly or by successive absorption/re-emission \citep{mckinneyDustenshrouded2021}, or from intense star-formation, similar to what is seen in local ULIRGS \citep[e.g.~Arp 220;][]{wilsonExtreme2014}.

%As shown in Figure~\ref{fig:stacking}, a moderately warm dust temperature $\sim 75$ K would be completely missed at the current limits of our samples. 

At the same time, the lack of mid-IR detections in LRDs appear at odds with the AGN interpretation. 
The one object in our sample individually-detected in MIRI/F1800W (COS-105829) exhibits a flat SED from $\sim 7$--$18$\,\textmu m, as does the median stack. 
Similar results have been reported for little red dots covered by the SMILES and JADES programs, with deep MIRI imaging in multiple bands \citep{williamsGalaxies2023,perez-gonzalezWhat2024}. 
As discussed in Section~\ref{sec:continuum}, this may indicate minimal contribution from hot dust in the torus, which in turn may imply that the AGN contribution is overestimated. 
However, an alternative explanation is that significant torus dust is not common in this sample, despite being ubiquitous in lower-redshift AGN.
Dust-deficient quasars are known to exist both in the local universe \citep{lyuDustdeficient2017,haoHotdustpoor2010} and at $z\sim 6$ \citep{jiangDustfree2010}, though these objects are typically blue in the rest-frame optical, lacking significant dust attenuation. 
Given that the LRDs exhibit significant attenuation, the dust deficit is rather a \textit{hot} dust deficit---a lack of dust at temperatures $\gtrsim 500$ K.

Perhaps a natural explanation is that the remarkable abundance and characteristic dust-reddening of the LRDs is directly related to their missing mid-IR emission. 
In the canonical picture of obscured AGN, the hot dust emitting in the mid-IR is also the material obscuring the AGN, with $A_V \gtrsim 10$ mag across the lines of sight blocked by the torus \citep{hickoxObscured2018}. 
In the LRDs, this obscuring material may instead be located in clouds further from the central source, with colder dust. 
These may originate from nuclear starburst disks \citep[e.g.][]{thompsonRadiation2005}, or from ``polar dust'' in the NLR, ejected via winds \citep[e.g.][]{lyuPolar2018}. 
The hottest dust, in the clouds at the sublimation radius, may itself be optically thick, or be heated inefficiently due to obscuration by the (dust-free) BLR clouds \citep[see][]{maiolinoJWST2024}.
Regardless, if the colder obscuring clouds have a high covering factor, i.e.~a roughly isotropic distribution, it could explain the ubiquity of reddened AGN at a given luminosity.

Such a geometry may be expected if these objects are in a nascent evolutionary stage. 
At $z<2$, red quasars are thought to be in the blowout phase; as the ``effective Eddington ratio'' is lower for dusty gas, as opposed to pure ionized Hydrogen, moderate column densities of dusty gas around the accretion disk can be easily blown out via radiative feedback \citep{fabianRadiative2006,fabianEffect2008a,ricciClose2017a,glikmanAccretion2024}.
In this case, the AGN effectively goes super-Eddington, blowing out its obscuring material, clearing lines of sight directly to the accretion disk. 
In addition to radiatively-driven outflows, MHD winds driven off the accretion disk may present the right conditions for dust formation, and may even be linked to the formation of the dust torus \citep{sarangiDust2019a}.
In fact, even the simplest models for torus formation begin with a larger, isotropic distribution of clouds, which form into a torus via anisotropic radiatively-driven winds \citep{liuDusty2011,bannikovaDynamics2017}.
The mid-IR deficit in the LRDs may therefore indicate that dynamic feedback/winds are at play, regulating the dust content at the sublimation radius, or that the torus has not fully formed at this epoch.

\subsection{Where there are AGN, there are galaxies}

Finally, we want to emphasize that despite the focus of this paper on the two extreme physical interpretations of the LRDs, neither is likely completely true. 
The physical mechanisms that trigger starbursts and AGN are similar, and we would expect that where there is early AGN growth, there is also significant star-formation \citep{thompsonRadiation2005,hopkinsCosmological2008,daviesStellar2009}. 
Already, we have clear evidence for AGN in the LRDs in the form of ubiquitous broad lines \citep{greeneUNCOVER2024}, and also evidence for evolved stellar populations with clear Balmer breaks \citep{wangRUBIES2024a}. 
Where these objects fit into the picture of black-hole galaxy co-evolution is yet to be determined.

It is widely believed that sub-millimeter galaxies at $z>2$ (i.e., 850 \textmu m-selected dusty star-forming galaxies) represent the progenitors of massive $z\sim 1$--$2$ elliptical galaxies \citep[e.g.][]{toftSubmillimeter2014, longMissing2023}. 
Stellar age gradients in quiescent galaxies imply an inside-out formation scenario, in which the central cores formed in early, rapid starbursts, consistent with the compact sizes of SMGs \citep{langBulge2014,suessColor2020}.

Extending this framework to higher redshift, the LRDs at $z\sim 5$--$9$ could represent the progenitors of at least some of the  first compact quiescent galaxies at $z\sim 3$--$4$. 
Recent \JWST\ observations find evidence for an over-abundance of massive quiescent galaxies at $z\sim 3$--$5$ \citep{carnallSurprising2023,valentinoAtlas2023}, including several which are quite compact \citep{itoSize2024,wrightRemarkably2024} or maximally old, with formation redshifts $\gtrsim 10$ \citep{glazebrookMassive2024, degraaffEfficient2024, carnallJWST2024}.
The overall volume density of the LRDs in our sample is $n\sim 2\times 10^{-5}$ Mpc$^{-3}$, similar to the overall volume density of massive ($\log M_\star/M_\odot > 10.6$) quiescent galaxies presented in \citet{valentinoAtlas2023}. 
Given also their compact sizes and similar formation redshifts, the LRDs may therefore be a natural progenitor population for these early quiescent galaxies. 
Simulations predict that AGN feedback drives quenching \citep[e.g.][]{daveMufasa2017, weinbergerSupermassive2018, kurinchi-vendhanOrigin2023}, and \citet{carnallMassive2023} find significant broad H$\alpha$ emission in the spectrum of an otherwise quiescent galaxy at $z=4.5$. 
However, a number of complicating factors muddy any conclusions regarding the connections between LRDs and quiescent galaxies. 
Samples of quiescent galaxies and LRDs are not mass-matched (nor are the stellar masses of LRDs well-constrained), and cover different redshift ranges. 
Moreover, the duty cycle of the LRD phase would have to be of order unity if they are the direct progenitors of massive quiescent galaxies. 
Understanding any potential evolutionary connection between LRDs and lower-redshift compact quiescent galaxies will require systematic follow-up and careful analysis to constrain the stellar contribution to the continuum.

\section{Conclusions}

We have presented a large sample of compact, extremely red objects/little red dots in COSMOS-Web. 

\begin{enumerate}
\item We present a sample of 434 LRDs selected based on their F444W compactness and F277W--F444W color $>1.5$, after removing likely brown dwarfs. By selection, we focus on the reddest subset of the LRD population, biasing our sample towards higher redshifts ($z\gtrsim 5$) but mitigating contamination from EELGs. 
\item We fit each candidate to both galaxy and quasar SED models to explore the two ``edge cases,'' or the two alternatively physical interpretations of the LRDs: either they are dominated by galaxy light from a compact, dusty starburst, or AGN light from a reddened quasar. 
\item We report the spectroscopic confirmation of one source, COS-756434, at $z=6.9993\pm 0.0001$, from public \JWST/NIRSpec PRISM data. The source exhibits clear broad lines in H$\alpha$ and H$\beta$, indicating strong AGN contribution. 
\item Under the interpretation of the LRDs as red quasars, we measure bolometric luminosities $\sim 10^{45-47}$ erg\,s$^{-1}$, spanning the gap between other \JWST-selected AGN and low-luminosity UV-selected quasars. We infer a bolometric luminosity function $\sim 100$ times higher than that for UV-selected quasars, consistent with lower limits from spectroscopic observations, implying a black hole accretion rate density at $z\sim 5$--$9$ rivaling that at cosmic noon. 
\item Under the alternative interpretation of the LRDs as massive/compact galaxies, perhaps hosting AGN but not dominated by their light, we measure stellar masses $\sim 10^{9-11}~M_\odot$. While many of the LRDs have masses too large to be found in this volume in a $\Lambda$CDM cosmology, when focusing on the subset of the sample with MIRI/F770W coverage, no source violates the limits imposed by $\Lambda$CDM. However, they dominate the high-mass end of the stellar mass function, and based on their compact sizes, must approach or exceed the maximal stellar mass surface densities observed in the local universe. 
\item We search all available X-ray, mid-IR, far-IR/sub-mm, and radio data in the COSMOS field to examine the multiwavelength properties of the LRDs. No source is individually detected, and stacks of the full sample yield non-detections in all bands except \HST/\JWST. We provide the median-stacked limits and a model for the maximal SED consistent with the stacked constraints. This maximal SED model requires an anomalously low X-ray normalization ($\alpha_{\rm OX} \lesssim -2$) or Compton-thick absorption ($\log N_{\rm H} \gtrsim 24.5$) and is radio-quiet ($R^*_{1.4} \lesssim 10$). The model also allows for a substantial warm dust component with $T_{\rm dust} \sim 200\pm 75$, K despite non-detections in the mid-IR and millimeter wavelengths. 
\end{enumerate}

The interpretation of the ``little red dots'' as a population of reddened AGN has important implications for our understanding of black hole growth in the early universe. 
Future large spectroscopic surveys will be needed to conduct a comprehensive census of LRDs, to elucidate their true nature at a population level, and deeper observations in the X-ray, mid/far-IR, and radio will be needed to determine their true properties.

\facilities{\textit{HST} (ACS), \textit{JWST} (NIRCam, NIRSpec, and MIRI)}

\software{\texttt{bagpipes} \citep{carnallInferring2018a}, \texttt{SEP} \citep{barbarySEP2016}, SourceExtractor \citep{bertinSExtractor1996}, \texttt{astropy} \citep{astropycollaborationAstropy2013}, \texttt{matplotlib} \citep{hunterMatplotlib2007}, \texttt{numpy} \citep{harrisArray2020}, \texttt{photutils} \citep{bradleyAstropy2022}, \texttt{scipy} \citep{virtanenSciPy2020}, STScI JWST Calibration Pipeline \citep[\url{jwst-pipeline.readthedocs.io};][]{rigbyScience2023}.}

\section*{Acknowledgements}

The authors thank Gene Leung, Ryan Endsley, Jenny Greene, and Ignas Juodžbalis for useful discussions on the nature of the ``little red dots.''
Support for this work was provided by NASA through grant JWST-GO-01727 and HST-AR-15802 awarded by the Space Telescope Science Institute, which is operated by the Association of Universities for Research in Astronomy, Inc., under NASA contract NAS 5-26555. 
H.B.A. acknowledges the support of the UT Austin Astronomy Department and the UT Austin College of Natural Sciences through Harrington Graduate Fellowship, as well as the National Science Foundation for support through the NSF Graduate Research Fellowship Program.
C.M.C. thanks the National Science Foundation for support through grants AST-1814034 and AST-2009577 as well as the University of Texas at Austin College of Natural Sciences for support; C.M.C. also acknowledges support from the Research Corporation for Science Advancement from a 2019 Cottrell Scholar Award sponsored by IF/THEN, an initiative of Lyda Hill Philanthropies. 
FG acknowledges the support from grant PRIN MIUR 2017-20173ML3WW\_001. 'Opening the ALMA window on the cosmic evolution of gas, stars, and supermassive black holes.'
H.B.A., C.M.C., and others at UT-Austin acknowledge that they work at an institution that sits on indigenous land. 
The Tonkawa lived in central Texas, and the Comanche and Apache moved through this area. 
We pay our respects to all the American Indian and Indigenous Peoples and communities who have been or have become a part of these lands and territories in Texas.

\newpage

\newpage
\bibliographystyle{aasjournal} 
\bibliography{cosmosweb_LRD_overview_paper.bib}

\newpage
\allauthors

\end{document}